\documentclass[12pt]{article}
\usepackage{graphicx,bm,amsmath,amssymb,latexsym,epsfig,euscript,multirow,color,url}
\usepackage[compress,numbers]{natbib}
\usepackage[colorlinks=true
,urlcolor=blue
,anchorcolor=blue
,citecolor=blue
,filecolor=blue
,linkcolor=blue
,menucolor=blue
,pagecolor=blue
]{hyperref}

\allowdisplaybreaks

\numberwithin{equation}{section}

\long\def\symbolfootnote[#1]#2{\begingroup%
\def\thefootnote{\fnsymbol{footnote}}\footnote[#1]{#2}\endgroup}

\newcommand{\be}{\begin{equation}}
\newcommand{\ee}{\end{equation}}
\newcommand{\pd}{\partial}
\newcommand{\bea}{\begin{eqnarray}}
\newcommand{\eea}{\end{eqnarray}}
\newcommand{\mat}{\begin{pmatrix}}
\newcommand{\rix}{\end{pmatrix}}
\newcommand{\nn}{\nonumber}
\renewcommand{\l}{\left}
\renewcommand{\r}{\right}

\renewcommand{\bar}{\overline}

\newcommand{\vv}{\mathbf}
\newcommand{\cO}{\mathcal{O}}
\newcommand{\cM}{\mathcal{M}}
\newcommand{\cL}{\mathcal{L}}
\newcommand{\BS}{\mathcal{B}}
\newcommand{\q}{\quad}
\newcommand{\qq}{\qquad}
\newcommand{\qqq}{\qquad\quad}
\newcommand{\qqqq}{\qquad\qquad}
\newcommand{\qqqqq}{\qquad\qquad\quad}

\newcommand{\comment}[1]{}

\input prepictex
\input pictex
\input postpictex
\newdimen\tdim
\tdim=\unitlength

\begin{document}

\begin{titlepage}

\begin{flushright}
\small{RUNHETC-2012-04}\\
\end{flushright}

\vspace{0.5cm}
\begin{center}
\Large\bf
Probing Colored Particles\\
with Photons, Leptons, and Jets
\end{center}

\vspace{0.2cm}
\begin{center}
{\sc Yevgeny Kats\symbolfootnote[1]{kats@physics.rutgers.edu} and Matthew J. Strassler\symbolfootnote[2]{strassler@physics.rutgers.edu}}\\
\vspace{0.6cm}
\textit{New High Energy Theory Center\\
Department of Physics and Astronomy\\
Rutgers University, Piscataway, NJ 08854, USA}
\end{center}

\vspace{0.5cm}
\begin{abstract}
\vspace{0.2cm}
\noindent

If pairs of new colored particles are produced at the Large Hadron Collider, determining their quantum numbers, and even discovering them, can be non-trivial. We suggest that valuable information can be obtained by measuring the resonant signals of their near-threshold QCD bound states. If the particles are charged, the resulting signatures include photons and leptons and are sufficiently rich for unambiguously determining their various quantum numbers, including the charge, color representation and spin, and obtaining a precise mass measurement. These signals provide well-motivated benchmark models for resonance searches in the dijet, photon+jet, diphoton and dilepton channels. While these measurements require that the lifetime of the new particles be not too short, the resulting limits, unlike those from direct searches for pair production above threshold, do not depend on the particles' decay modes. These limits may be competitive with more direct searches if the particles decay in an obscure way.

\end{abstract}
\vfil

\end{titlepage}

\tableofcontents

\newpage
\section{Introduction and Motivation}

The high collision energy available at the Large Hadron Collider (LHC) permits the exploration of a vast region of particle physics territory that was
previous inaccessible.  To assure sensitivity to the rich array of signatures that may arise from physics beyond the Standard Model, a diverse set of analysis techniques is needed.  Already, limits on some particles have reached as high as 2 TeV and beyond.  But in some cases, the limits are much weaker.  New colored particles even as light as 200~GeV, which would have large production cross sections, are still not excluded,  because of difficult backgrounds, trigger limitations, or the need for dedicated analysis methods.

In this paper we will show that pair-produced colored particles from many Beyond-the-Standard-Model scenarios are constrained, in a way that is largely model-independent, by the non-observation of dijet (or other) resonances arising from their QCD bound states. In some cases the constraints obtained by this method would be stronger than the ones from the more standard searches. Furthermore, if new pair-produced colored particles are discovered, by any method, their bound state signals can provide a uniquely powerful tool for determining their quantum numbers, including charge, spin, color representation and multiplicity.  With such quantum numbers determined, the bound state resonances then also provide a high-precision mass measurement.  This approach is especially useful for particles in color representations higher than the triplet (so that the bound states have a sizeable cross-section) that also carry electric charge (so that multiple resonant signals may be observable, including $\gamma\gamma$, $\gamma$+jet and $\ell^+\ell^-$).

These ideas are hardly new, of course.  Bound states were crucial in understanding charm and bottom quarks, and had the top quark been lighter than 130 GeV, measurements of toponium at $e^+e^-$ and hadron colliders (see, for example,~\cite{Fadin:1987wz,Kuhn:1987ty,Strassler:1990nw,Pancheri:1992km}) could have provided very detailed information about its properties.  But it has been some time since these ideas were current, and so they need to be dusted off and shined for use at the LHC.  This has been done recently in~\cite{Kats:2009bv} and further in~\cite{Kahawala:2011pc}, where the focus was on bound states of particles that appear in theories of supersymmetry or extra dimensions, though other cases were briefly discussed. Utilizing bound state annihilation signals at the LHC has been considered also in~\cite{Drees:1993yr,Drees:1993uw,Chikovani:1996bk,Arik:2002nd,Cheung:2004ad,BouhovaThacker:2004nh,BouhovaThacker:2006pj,Bussey:2006vx,Martin:2008sv,Kim:2008bx,Kauth:2009ud,Barger:2011jt}. In this work we explore these ideas in a much more general context and confront the potential signals with data from the 7~TeV LHC.

The only essential model-dependent assumption that we make about the new particle $X$ is that its life is slightly prolonged, so that the $X\bar X$ bound state (``$X$-onium'') decays not by the disintegration of the $X$ itself (as in toponium) but by the annihilation of the $X$ with the $\bar X$ (as in charmonium). This condition is satisfied relatively easily as long as the particle does not have any unsuppressed 2-body decays. For particles in higher representations of color this situation becomes even more generic because, as we will see, the annihilation rates are proportional to many powers of color factors and can be enhanced by orders of magnitude.

Particles in the triplet and octet representations of color appear in many common scenarios and our results will be relevant to many of them. We will also consider particles in more exotic representations, for some of which the signals we discuss can be unusually large.   (Other aspects of the phenomenology of massive vectorlike matter in exotic representations have been considered in the past in~\cite{Wilczek:1976qi,Dover:1979sn,Lust:1985aw,Borisov:1986ev,Chivukula:1990di,Chivukula:1991zk,DelNobile:2009st,delAguila:2010mx,Han:2010rf,Kumar:2011tj}.)  Our motivations for discussing such representations are several.  First, searches for resonances will be done anyway, and can be interpreted in terms of any representations, so it makes sense to understand their implications for all available representations without imposing theoretical assumptions. Second, massive vector-like matter in other representations often arises both in perturbative string theory, our only top-down model framework, and quantum field theory models with strong dynamics (or their dual warped-extra-dimensional cousins). In fact, string theory typically predicts extra matter that, though chiral at high energies, becomes vector-like after symmetry breaking which may well occur at or just above the weak scale.  The types of matter commonly predicted by string theory are biased by the fact that strings have two ends, and each end carries an index under a gauge or flavor group.  This means that matter in perturbative string theory tends to appear in a limited set of representations with two indices, namely adjoints, symmetric and antisymmetric tensors, bifundamentals, and fundamentals.  At the least, therefore, one should consider the possibility of massive sextet representations of color.  Meanwhile, if non-perturbative quantum field theory has a role to play near the weak scale, composite particles in more complicated representations may arise through new confining dynamics.  In QCD we encounter massive octet and decuplet states of flavor-$SU(3)$ (with $U(1)$-baryon charge) around the scale of confinement; in a similar vein, we might easily find technibaryons or other confined states near the TeV scale in the octet or decuplet of color, and carrying $U(1)$ hypercharge.   Furthermore, the two-index tensor representations expected in perturbative string theory might be combined through strong dynamics into more elaborate massive bound states.  This can occur in confining supersymmetric quantum field theories, where massless states with high representations of global symmetries (into which $SU(3)$-color may be embedded) sometimes arise.

In section~\ref{sec-prelim} we will set the stage for our study by discussing the various colored particles that we will consider and the degree to which they are constrained by current data. In section~\ref{sec-BS} we will review the basics of the bound state formalism. In section~\ref{sec-signals} we will analyze the bound state annihilation signals in the $\gamma\gamma$, $\gamma$+jet, dijet and $\ell^+\ell^-$ channels. In section~\ref{sec-widths} we will comment on the widths of the resonances. In section~\ref{sec-strategy} we will present a strategy that combines the information from the various channels for determining the properties of the new particles. We will summarize and give final remarks in section~\ref{sec-summary}.

\section{New Colored Particles: Preliminary Considerations}
\label{sec-prelim}

Since the properties of the bound state signals are commonly model-independent, we will not restrict ourselves to a particular new physics scenario. We will simply assume that nature contains a new pair-produced particle $X$ of spin $j=0$, $\frac12$ or $1$, electric charge $Q$, and color representation $R$ (details of various color representations are given in table~\ref{tab-reps}).  For simplicity and brevity, we will assume that $X$ is a singlet of the electroweak $SU(2)$, though our methods apply also to other $SU(2)$ representations. We also assume that the only interactions that contribute to production and annihilation of $X\bar X$ pairs, and the $X\bar X$ binding potential, are the Standard Model gauge interactions.\footnote{Our results can be significantly modified for particles which have large Yukawa-type interactions, such as gluinos (when the squarks are not much heavier)~\cite{Kats:2009bv,Kahawala:2011pc} or fourth-generation quarks~\cite{Barger:1987xg}.  Note, however, that some of the representations that we study here cannot have any renormalizable non-gauge interactions with Standard Model particles, simply because their quantum numbers do not allow it.  Even when other renormalizable couplings are permitted, any subsequent effects are competing against the large value of $\alpha_s$, and so our formulas for production and decay rates will often be accurate even in this case.} Then, at least in the case of $j=0$ or $\frac12$ particles, everything is fixed by gauge invariance. The $j=1$ case is a bit more subtle and discussed in appendix~\ref{app-spin1}.

\begin{table}[t]
$$\begin{array}{|c|c|c|c|c|c|c|}\hline
R     &\, D_R   & C_R   &  T_R & A_R &\,t_R & \; R \otimes \bar R \; \\\hline\hline
(1,0) & \vv{3}  & 4/3   &  1/2 &  1  &   1  & \vv{1} \l(\oplus \vv{8}\r) \\\hline
(1,1) & \vv{8}  & 3     &    3 &  0  &   0  & \vv{1} \oplus \vv{8} \oplus \vv{8} \l(\oplus \vv{10} \oplus \vv{\bar{10}} \oplus \vv{27} \r) \\\hline
(2,0) & \vv{6}  & 10/3  &  5/2 &  7  &   2  & \vv{1} \oplus \vv{8} \l(\oplus \vv{27}\r) \\\hline
(2,1) & \vv{15} & 16/3  &   10 & 14  &   1  &\,\vv{1} \oplus \vv{8} \oplus \vv{8} \oplus \vv{10} \oplus \vv{\bar{10}} \oplus \vv{27} \oplus \vv{27} \l(\oplus \vv{35} \oplus \vv{\bar{35}} \oplus \vv{64}\r) \\\hline
(3,0) & \vv{10} & 6     & 15/2 & 27  &   0  & \vv{1} \oplus \vv{8} \oplus \vv{27} \l(\oplus \vv{64}\r) \\\hline
(2,2) & \vv{27} & 8     &   27 &  0  &   0  &
\begin{array}{c} \vv{1} \oplus 2\cdot\vv{8} \oplus \vv{10} \oplus \vv{\bar{10}} \oplus 3\cdot\vv{27} \oplus 2\cdot\vv{35} \oplus 2\cdot\vv{\bar{35}} \oplus  2\cdot\vv{64} \\ \l(\oplus \vv{28} \oplus \vv{\bar{28}} \oplus \vv{81} \oplus \vv{\bar{81}} \oplus \vv{125}\r) \end{array}\\\hline
(3,1) & \vv{24} & 25/3  &   25 & 64  &   2  &
\begin{array}{c} \vv{1} \oplus 2\cdot\vv{8} \oplus \vv{10} \oplus \vv{\bar{10}} \oplus 2\cdot\vv{27} \oplus \vv{35} \oplus \vv{\bar{35}} \oplus  2\cdot\vv{64} \\ \l(\oplus \vv{81} \oplus \vv{\bar{81}} \oplus \vv{125}\r) \end{array} \\\hline
(4,0) & \vv{15'}& 28/3  & 35/2 & 77  &   1  & \vv{1} \oplus \vv{8} \oplus \vv{27} \oplus \vv{64} \l(\oplus \vv{125}\r) \\\hline
\end{array}$$
\caption{Properties of $SU(3)$ representations~\cite{Slansky:1981yr,Cutler:1999vc,Wesslen:2009} ordered by their quadratic Casimir $C_R$. Also shown are the dimension $D_R$, the index $T_R$, the anomaly coefficient $A_R$, and the triality $t_R$. The last column shows the representations obtained by combining the representation $R$ with its complex conjugate (where representations that give rise to non-binding potential between the particles are enclosed in parentheses). Our conventions and useful $SU(3)$ identities are given in appendix~\ref{app-group-theory}.}
\label{tab-reps}
\end{table}

\subsection{Constraints on quantum numbers}
\label{sec-QN}

New stable colored particles would form exotic atoms that are not observed~\cite{Hemmick:1989ns}, so we will focus on particles $X$ that are able to decay down to known particles, possibly along with invisible new particles. This restricts the charge $Q$ to be an integer times $1/3$ and imposes a correlation between $Q$ and the triality $t_R$ of $R$ [which can be computed from~(\ref{triality})].  For any representation $R$ in the decomposition of $R_1\otimes R_2$,
\be
t_R = \l(t_{R_1} + t_{R_2}\r)\,\mbox{mod}\,3
\ee
and thus triality is conserved mod~3.  Quarks ($R={\bf 3}$; $Q=-1/3$ or $+2/3$), antiquarks ($R={\bf\bar 3}$; $Q=+1/3$ or $-2/3$) and gluons ($R={\bf 8}$; $Q=0$) all have charge and triality that satisfy
\be
t_R = (-3Q)\,\mbox{mod}\,3
\label{tR-Q}
\ee
as do the colorless particles of the Standard Model, and any new invisible particles.  Consequently, for $X$ to be able to decay, it must also satisfy~(\ref{tR-Q}).   Particles $X$ in representations with vanishing triality (including $\vv{8}$, $\vv{10}$, and $\vv{27}$) therefore must have integral charges, while particles in other representations (including $\vv{3}$, $\vv{6}$, $\vv{15}$, $\vv{24}$, $\vv{15'}$) must have fractional charges which are multiples of $1/3$.  Spin $j$ is unconstrained by these considerations, since three quarks can have the quantum numbers of a neutron ($j=1/2$; $R={\bf 1}$; $Q=0$) allowing a change of spin by a half-integer with no change in charge or color.

One might consider constraining the list of particles by requiring that the running of the Standard Model gauge couplings should not be too severely affected. High representations in $j$ and $R$ can have a huge impact on the running of $\alpha_s$, and (for $Q\neq 0$) electroweak couplings.  Requiring asymptotic freedom in $SU(3)$ would preclude some of the representations we consider.  However, if $X$ were composite, then above its confinement scale the beta functions would change again, and asymptotic freedom might be restored.  Alternatively, in a sufficiently exotic scenario with a whole set of new particles, including vectors as well as scalars or fermions, their effects on the beta function may partially cancel.  Similar remarks apply to the potential contributions of the particles to loops affecting precision measurements. Lacking a clear-cut criterion, we leave these considerations to the reader's judgement, and impose no constraint of our own.

\subsection{Restrictions from collider searches}

Let us now consider the direct constraints on new colored particles from various collider measurements. These may be especially stringent for particles in high representations in $j$ and $R$, which have large pair-production cross-sections,\footnote{Since we assume $X$ has a small width, resonant production of $X$ is too rare to be important.} as shown in figures~\ref{fig-cs-pairs-Tevatron}--\ref{fig-cs-pairs-LHC} (based on the expressions in appendix~\ref{app-pair-xsec}).\footnote{We assumed the fields $X$ to be complex. Electrically-neutral fields (of any spin) in the $\vv{8}$ or $\vv{27}$ can be real, in which case the cross sections would be smaller by a factor of $2$.}  However, existing limits depend very significantly on the particles' decay channels, while our aim is to remain as model-independent as possible. We will discuss several examples but will be cautious in applying constraints except where obviously necessary.

\begin{figure}[t]
\begin{center}
\includegraphics[width=0.91\textwidth]{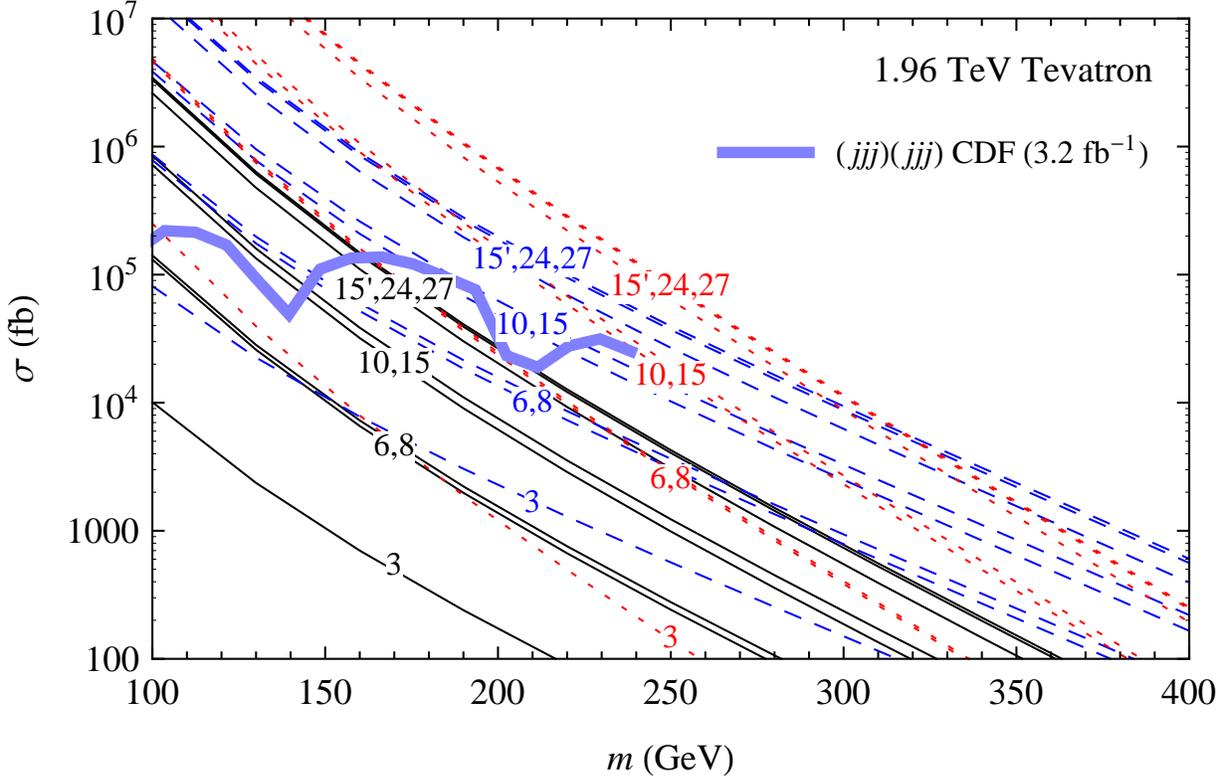}
\caption{Tevatron pair-production cross sections for particles of mass $m$ and spin $j=0$ (solid black), $\frac{1}{2}$ (dashed blue) or $1$ (dotted red) in color representations $R$ indicated in the figure. The 95\% CL exclusion limit on particles decaying to 3 jets is the thick blue curve (CDF~\cite{:2011sg,CDF-10256,Seitz:2011zz}, $3.2$~fb$^{-1}$).}
\label{fig-cs-pairs-Tevatron}
\end{center}
\end{figure}

\begin{figure}[t]
\begin{center}
\includegraphics[width=0.91\textwidth]{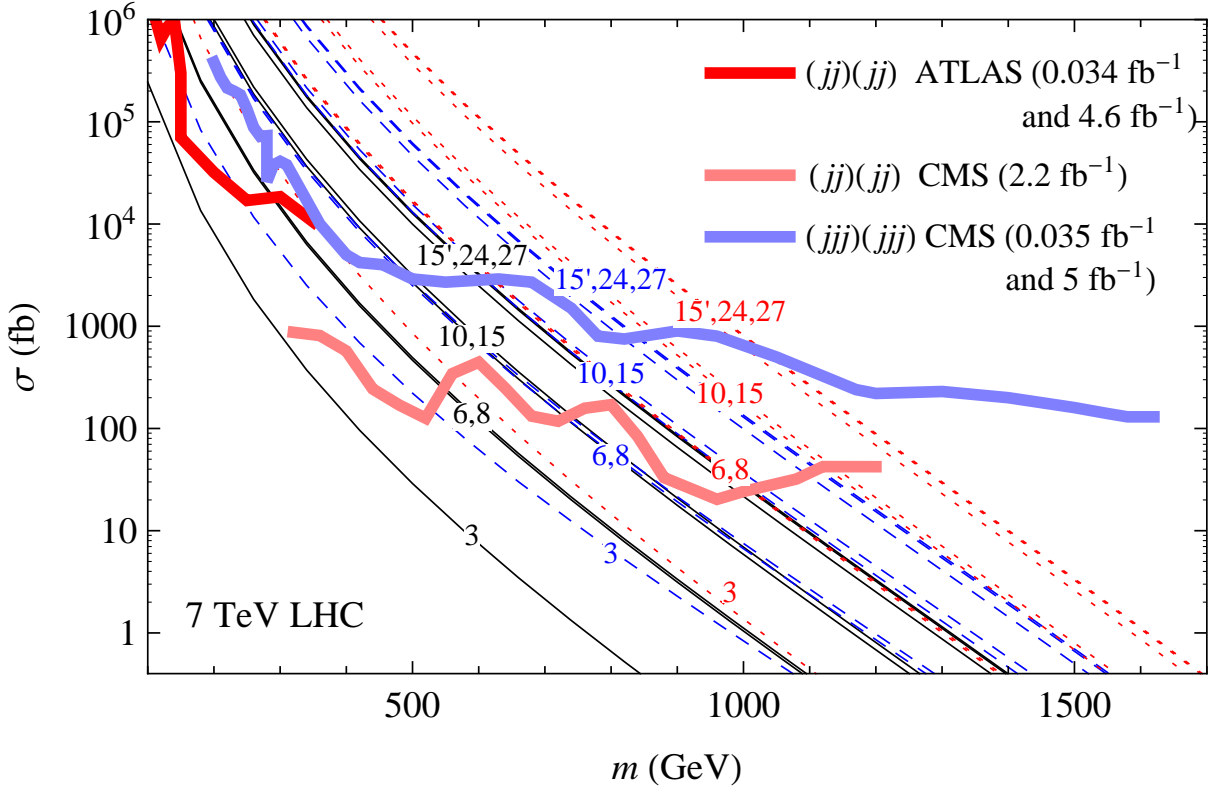}
\caption{7~TeV LHC pair production cross sections for particles of mass $m$ and spin $j=0$ (solid black), $\frac{1}{2}$ (dashed blue) or $1$ (dotted red) in color representations $R$ indicated in the figure. The 95\% CL exclusion limits on particles decaying to 2 jets are the thick red curve (ATLAS with 4.6~fb$^{-1}$~\cite{ATLAS-CONF-2012-110} for $m > 150$~GeV, 34~pb$^{-1}$~\cite{Aad:2011yh} for lower masses) and the thick pink curve (CMS with 2.2~fb$^{-1}$~\cite{CMS-PAS-EXO-11-016}). The limit on particles decaying to 3 jets is the thick blue curve (CMS with 5~fb$^{-1}$~\cite{CMS-3jet,Duggan-talk} for $m > 280$~GeV, 35~pb$^{-1}$~\cite{Chatrchyan:2011cj} for lower masses).}
\label{fig-cs-pairs-LHC}
\end{center}
\end{figure}

\begin{table}[t]
$$\begin{array}{|c||c|c|c|c|c|c|c|c|}\hline
              &\vv{3} &\vv{8} &\vv{6} &\vv{15}& \vv{10}  & \vv{27}  &\vv{24}&\vv{15'}\\\hline\hline
\;j = 0     \;&\q 2 \q&\q 2 \q&\q 2 \q&\q 3 \q&\;2\,(3)\;&\;2\,(3)\;&\q 3 \q&\q 4 \q\\\hline
  j=\frac12   &   2   &   3   &   2   &   2   &     3    &    4     &   3   &   3  \\\hline
  j = 1       &   2   &   2   &   2   &   3   &     3    &    3     &   3   &   4  \\\hline
\end{array}$$
\caption{The minimal number of jets to which a spin-$j$ particle in a particular color representation can decay, accounting for color, charge and angular momentum conservation.  As discussed in the text, the $\vv{8}$, $\vv{10}$ and $\vv{27}$ must have integer electric charges while the others' must be fractional.  For particular values of the charge, the minimal number of jets will sometimes be higher than shown in the table.  For the $\vv{10}$ or $\vv{27}$, $j=0$, any non-zero charge requires the particle to decay to at least three jets, as indicated in parentheses.}
\label{tab-Njets}
\end{table}

One signature particularly relevant to particles in high representations of color is decays to jets only. In table~\ref{tab-Njets} is shown the minimal number of quarks, antiquarks and/or gluons to which a particle $X$ with spin $j$ and color representation $R$ could decay, consistent with conservation of color, angular momentum and charge.  Notice that a high-dimensional operator is sometimes required for the corresponding decay, automatically suppressing decays with easier signatures (except possibly for decays involving top quarks), and at the same time giving a sufficiently long lifetime for bound states to annihilate.\footnote{If the lifetime of $X$ is \emph{very} large, then when produced above threshold it may travel a macroscopic distance in the detector (within an R-hadron).  While our results would still apply, the direct observation of such particles would typically make discovery and measurement of quantum numbers much easier (see, e.g.,~\cite{Graham:2011ah} and references therein) and constraints on them are already very powerful~\cite{CMS-PAS-EXO-11-022,ATLAS-CONF-2012-022}.} Note that this is neither the best nor the worst case as far as sensitivity to the new particles goes.  For instance, if the final state contains invisible particles and consequently a lot of missing transverse momentum, discovery would likely have occurred through searches aimed at supersymmetry, while if the new particles decay to jets plus a new colorless particle that itself typically decays to three jets, it is not obvious whether any searches would be sensitive other than those seeking black holes and similar phenomena.

Signatures with jets only, and no missing energy, have been traditionally considered hard due to the large QCD background. Nevertheless, several successful search techniques for pair-produced hadronically decaying particles have been developed recently, starting with searches for particles decaying to 3 jets by CDF~\cite{:2011sg,CDF-10256,Seitz:2011zz} ($3.2$~fb$^{-1}$) and CMS~\cite{Chatrchyan:2011cj} ($35$~pb$^{-1}$, recently updated to $5$~fb$^{-1}$~\cite{CMS-3jet}\footnote{The article~\cite{CMS-3jet} presents the limit curve only up to $m = 1000$~GeV, but a more complete curve (up to 1600~GeV) has been presented in~\cite{Duggan-talk}.}), and followed by searches for particles decaying to 2 jets by ATLAS~\cite{Aad:2011yh} ($34$~pb$^{-1}$, recently updated to $4.6$~fb$^{-1}$~\cite{ATLAS-CONF-2012-110}) and CMS~\cite{CMS-PAS-EXO-11-016} ($2.2$~fb$^{-1}$). Estimates of their 95\% CL exclusion limits are shown\footnote{For presenting the results of~\cite{CMS-PAS-EXO-11-016}, we assumed the acceptance to be the same as for the coloron examined in~\cite{CMS-PAS-EXO-11-016}. We determined the acceptance by dividing the acceptance-times-cross-section curve from~\cite{CMS-PAS-EXO-11-016} by the cross section of the model used in that study~\cite{Dobrescu}. This is consistent with the two values quoted in~\cite{CMS-PAS-EXO-11-016} ($3\%$ for $m = 300$~GeV and $13\%$ for $m = 1000$~GeV).} as thick curves in figures~\ref{fig-cs-pairs-Tevatron}--\ref{fig-cs-pairs-LHC}. These limits are likely to be conservative since the cross sections we plot do not include higher-order QCD corrections (which for many cases are not yet available in the literature) which would typically make them bigger.\footnote{One correction that will be especially important for the high color representations is the Sommerfeld enhancement which, like the bound states, originates from the attractive QCD potential between the two particles (see, e.g.,~\cite{Fadin:1990wx}). We have estimated that typically (e.g., for 1~TeV particles at the 7~TeV LHC or 200~GeV particles at the Tevatron) the enhancement factors relative to the leading-order cross sections would be $\sim 1$-$2$ for $X$ in the $\vv{6}$ and $\vv{8}$ representations, $\sim 3$ for $\vv{10}$ and $\vv{15}$, and $\sim 5$ for $\vv{27}$, $\vv{24}$ and $\vv{15'}$. However, since the searches for hadronically decaying particles typically require the particles to be somewhat boosted, so as to reduce QCD multijet background, they are not sensitive to pairs produced near the threshold, where the effect of the Sommerfeld enhancement is most pronounced. As a result, the number of events passing the cuts will be enhanced by smaller factors than those stated above. Addressing this in more detail requires a separate study.} Meanwhile, the limits on spin-1 particles are conservative also, because we did not include the contribution from the $q\bar q$ channel in that case since it is necessarily model-dependent and typically subdominant, as discussed in more detail in appendix~\ref{app-spin1}. We should also note that the limits inferred from figures~\ref{fig-cs-pairs-Tevatron}--\ref{fig-cs-pairs-LHC} do not take into account that the angular and invariant mass distributions of the pairs depend on the spin and color representation, so the acceptance varies somewhat between the different cases.

For some of the representations the cross sections are so large that they are likely to be highly constrained even without dedicated searches, by the absence of anomalies in high-multiplicity events. Those have been explored recently in the CMS black holes search~\cite{Collaboration:2012ta} (with 4.7~fb$^{-1}$). While a careful interpretation of the limits from this search would require a detailed analysis, an examination of the backgrounds suggests that such an analysis would likely exclude certain regions, roughly within the 1000--1500~GeV mass range, for all the particles from table~\ref{tab-Njets} decaying to 4 jets (spin-$\frac12$ particles in the $\vv{27}$, and spin-0 and 1 particles in the $\vv{15'}$) and for some of the particles decaying to 3 jets (spin-1 particles in the $\vv{15}$, $\vv{10}$, $\vv{27}$, and $\vv{24}$, and spin-$\frac12$ particles in the $\vv{24}$ and $\vv{15'}$). Note from figure~\ref{fig-cs-pairs-LHC} that the resulting limits on these 3-jet cases would be complementary to those from the dedicated search for particles that decay this way. The dedicated 3-jet search~\cite{CMS-3jet,Duggan-talk}, which is motivated by models with low signal-to-background ratio, needs to use hard cuts for reducing the background, resulting in signal acceptance of 2-3\%. However, for signals that are larger than the background, or at least comparable to the uncertainty on the background, such cuts are no longer needed and limits can be set by inclusive searches like~\cite{Collaboration:2012ta} even when the number of signal events is small.

Also, for sufficiently large representations, the new particles will alter the overall dijet spectrum through large loop corrections. But the most prominent effect of these loops is likely to be below threshold, where they are enhanced by Coulomb effects and manifest themselves through the production and annihilation of bound states. Since the contribution from the bound states is resonant, and benefits from higher parton luminosity, it is probably the source of the most powerful limits arising from the dijet spectrum.  Limits of this type will be discussed below.

Overall, we see that the direct limits on hadronically decaying particles are not extremely strong, sometimes even for particles in high representations, and apply only in certain mass ranges for each specific decay mode. We will see that bound state signals may sometimes set limits of comparable strength which are less model-dependent. We also note that the direct searches for particles decaying to 3 jets suffer from uncertainties due to the very hard cuts needed for reducing the QCD background, leading to strong sensitivity to the far tails of the kinematic distributions which are difficult to simulate reliably.\footnote{For example, the cross section of a hadronically decaying top which can be inferred from the CDF search for particles decaying to 3 jets~\cite{:2011sg,CDF-10256,Seitz:2011zz} appears larger than expectation by a factor of $\sim 10$. While the excess can be accounted for by a $2\sigma$ fluctuation or indicate a new physics source of boosted top quarks, the discrepancy may also be attributed to underestimating the systematic uncertainty on the very small signal efficiency ($\sim 10^{-4}$).}  By contrast, the bound state signals have $\cO(1)$ efficiencies.

\section{Bound State Formalism}
\label{sec-BS}

Any new colored particle will produce a rich spectrum of bound states, similarly to the charmonium system. Often there will be even more variety, since particles in any representation other than the triplet can form bound states in more than one representation of color, and spin-1 particles provide additional possibilities for the spins of the bound states. However, for each of our signals, just one or a few of the bound states will contribute significantly to observable LHC signals. In particular, the probability to produce a bound state depends on the size of its wavefunction at the origin, $\psi(\vv{0})$, which is non-vanishing only for S waves. For any particular annihilation channel, the spins and the color representations of the bound states that need to be considered are further constrained. Luckily, as we will see, in most cases the contribution from the relevant S waves is not suppressed in any way, so we will not need to take into account higher angular momentum states or radiative transitions between states. Only the dilepton channel will require a more detailed analysis. For simplicity, in the rest of this introductory section we will focus on the case of directly produced S-wave bound states, and discuss the additional states relevant to the dilepton channel when we come to it. We will start by describing the leading-order expressions that we will be using and then comment on higher-order corrections, which become more and more important as we go higher in the color representation $R$.

\subsection{Leading order}

For particles of mass $m \gg \Lambda_{\rm QCD}$, and as long as the Bohr radius of the relevant bound state is much smaller than the QCD scale and the velocity of its constituents is non-relativistic, we can obtain reasonable estimates using a standard modified-hydrogenic approximation.  For a particle $X$ in representation $R$, the potential between $X$ and $\bar X$ depends on the color representation ${\cal R}$  of the $X\bar X$ pair through the quadratic Casimirs of $R$ and ${\cal R}$ as
\be
V(r) = -C\frac{\bar\alpha_s}{r}\,, \qqq
C = C_R - \frac12 C_{{\cal R}}
\label{potential}
\ee
Here
\be
\bar\alpha_s \equiv \alpha_s(r_{\rm rms})
\label{alphas-bar}
\ee
is defined as the running coupling at the scale of the average distance between the two particles in the corresponding hydrogenic state, $r_{\rm rms} \equiv \sqrt{\langle r^2\rangle}$, which is of the order of the Bohr radius\footnote{More precisely, $r_{\rm rms} = \sqrt{3}\,a_0$ for the S-wave ground state and $\sqrt{30}\,a_0$ for the lowest P-wave state. Since $a_0$ itself depends on $\bar\alpha_s$, we determine $r_{\rm rms}$ numerically using a self-consistency condition.\label{Bohr-radius}} $a_0 = 2/(C\bar\alpha_s m)$.  (The symbol $\alpha_s$ without a bar will be reserved for its value at the scale $m$.) The binding energies and the wavefunctions at the origin for the ground state $(n=1)$ and its radial excitations ($n=2,3,\ldots$) are given by
\be
E_b = -\frac{1}{4n^2}\, C^2\bar\alpha_s^2 m \,,\qq
\l|\psi(\vv{0})\r|^2 \equiv \frac{1}{4\pi}\l|R(0)\r|^2 = \frac{C^3 \bar\alpha_s^3 m^3}{8\pi\,n^3}
\label{Eb-wf}
\ee
and the cross-section for the bound state $\BS$ to be produced by initial-state partons $a$ and $b$ is
\be
\hat\sigma_{ab\to\BS}(\hat s)
= \frac{8\pi}{m}\,\frac{\hat\sigma_{ab\to X\bar X}^{\rm free}(\hat s)}{\beta(\hat s)}\, |\psi(\vv{0})|^2\,2\pi\,\delta(\hat s-M^2)
\label{sigma-hat-general}
\ee
where $M = 2m + E_b$ is the mass of the bound state, $\hat\sigma_{ab\to X\bar X}^{\rm free}(\hat s)$ is the production cross section for a free pair at threshold (i.e., for $\beta(\hat s) \to 0$, where $\beta(\hat s)$ is the velocity of $X$ or $\bar X$ in their center of mass frame).  (If $X$ is real, then $\bar X=X$, and eq.~(\ref{sigma-hat-general}) still holds with $\psi(\vv{0})$ defined through the expression~(\ref{Eb-wf}), rather than by the appropriately symmetrized wave-function.)
Meanwhile one may show that the production cross section of any narrow resonance $\BS$ of mass $M$ and spin $J$ from $a$ and $b$, and the decay rate back to $a$ and $b$, are related by
\be
\hat\sigma_{ab\to \BS}(\hat s) = \frac{2\pi \l(2J+1\r) D_{\BS}}{D_a D_b}\,
\frac{\Gamma_{\BS\to ab}}{M}\;2\pi\,\delta(\hat s - M^2)\qq \l(\,\times 2\;\mbox{ for } a = b \,\r)
\label{RelateSigmaToGamma}
\ee
where $D_p$ denotes the dimension of the color representation of particle $p$.

From (\ref{Eb-wf}), we may see that the binding energies would typically be small relative to the bound state mass. Even for high representations like the $\vv{15}$ and $\vv{10}$, the binding energy for a color-singlet S-wave bound state in its ground state is only $E_b \sim -0.05 M$. It is even smaller for bound states with non-zero orbital angular momentum and/or color. Therefore observing the splittings between the various states would generally be difficult at the LHC due to resolution limitations, except in the case of very high representations. For $R = \vv{10}$ and $\vv{15}$, observation of radially excited states distinctly from the ground state might barely be possible in diphoton and dielectron channels, but the rates for the excited states are very low and a very large data set would be required. Furthermore, annihilation rates fall like $1/n^3$, making radially excited states more vulnerable to decays of the constituent $X$ particles themselves, which could eliminate their annihilation signals.  For these reasons, we will assume in this paper that no information can be gleaned from the presence of several distinct states in the same channel.

When combined with detector resolution, the effect of the radially excited states, whose production rate falls as $1/n^3$, will be a small distortion on the high-energy side of the observed resonance shape, increasing the area under the observed resonance by only
\be
\zeta(3)=\sum_{n=1}^\infty \frac1{n^3} \approx 1.2
\label{zeta}
\ee
Since this effect is small and, as mentioned above, the radial excitations are more vulnerable to constituent decays, we will take the conservative approach of not including the factor (\ref{zeta}) in the cross sections.

\subsection{The importance of higher order corrections}
\label{sec-higher-order-intro}

The computations in this paper will all be at the leading order both in the short-distance production and annihilation amplitudes and in the non-relativistic and Coulomb-like treatment of the bound states,\footnote{We note that higher order results have been computed in the context of supersymmetric theories for bound states of color-triplet scalars (squarks)~\cite{Martin:2009dj,Younkin:2009zn} and color-octet fermions (gluinos)~\cite{Hagiwara:2009hq,Kauth:2009ud}.} with the exception of the running of $\alpha_s$ which we do account for to a degree, as described above, and of our use of NLO parton distribution functions (PDFs). We must therefore emphasize that our results will be rather imprecise. Our aim in this paper is not high precision, but rather to show that bound state searches may serve as a useful tool at the LHC, and to motivate future calculations of greater precision that will allow maximal information to be extracted from those searches. We will also demonstrate, however, that high precision is often not required for determining discrete quantum numbers.

The leading order approximation becomes less valid as we go up in the representations. In particular, the particle velocity in the ground state of a color-singlet S-wave bound state is $v \approx C_R\bar\alpha_s/2$, where $C_R$ is the second Casimir of the representation $R$, and thus for too high representations the bound states cannot be treated non-relativistically. The relativistic corrections contribute at the next-to-next-to-leading order in the $\bar\alpha_s$ expansion. Other corrections that depend on the combination $C_R\bar\alpha_s$ also start appearing at that order. For more details, see~\cite{Kauth:2009ud}, which studied the effect of such corrections on a color-singlet S-wave bound state of $R=\vv{8}$ fermions (gluinos). In their case, the corrections make $|\psi(\vv{0})|^2$ larger than the leading-order expression by a factor that varies from $1.8$ for $m=300$~GeV to $1.5$ for $m=1500$~GeV. Based on the values of $C_R$ from table~\ref{tab-reps}, we expect a comparable effect for $R = \vv{6}$ and a larger effect for $R = \vv{15}$ and $\vv{10}$ (where $C_R$ is about twice as large). For $R = \vv{27}$, $\vv{24}$ and $\vv{15'}$, $C_R\bar\alpha_s \approx 1$, so we cannot obtain reliable results for these (or higher) representations. Therefore in the following sections we will restrict our attention to $R = \vv{3}$, $\vv{8}$, $\vv{6}$, $\vv{15}$ and $\vv{10}$. We will briefly describe what the leading-order expressions give for higher representations in section~\ref{sec-high-reps-results}.

\section{Signals}
\label{sec-signals}

In this section we will identify the channels in which the bound state resonances would be most easily measurable and compute the corresponding cross sections as a function of the mass, color representation, charge and spin of the constituent particles.

The strong and electroweak couplings of the particles allow them to annihilate to a pair of gauge bosons, each of which can be a gluon, photon or $Z$. They can also annihilate to a pair of quarks through an $s$-channel gluon or any pair of fermions through an $s$-channel $\gamma/Z$. Since for simplicity we assumed the particles to be $SU(2)$ singlets, there are no processes involving $W$ bosons. The promising final states that we will analyze are diphoton ($\gamma\gamma$), photon+jet ($\gamma g$), dijet ($gg$ and $q\bar q$), and dilepton ($\ell^+\ell^-$).

Annihilation to $\gamma\gamma$, $\gamma g$ or $gg$ is possible only for bound states with total angular momentum $J = 0$ or $2$. Such bound states (color-singlet for $\gamma\gamma$, color-octet for $\gamma g$, and various representations for $gg$) can only be produced in the $gg$ channel.

Annihilation to $q\bar q$ or $\ell^+\ell^-$ (through an $s$-channel gauge boson) is possible only for $J = 1$ bound states. Color-octet $J = 1$ bound states which will annihilate to $q\bar q$ will be produced in the $q\bar q$ channel. On the other hand, there is no leading-order strong-interaction process that can produce the color-singlet $J = 1$ bound states which can annihilate to $\ell^+\ell^-$. Therefore, we will have to take into account various subleading processes which will lead to a measurable $\ell^+\ell^-$ signal.

The photon+jet, diphoton and dilepton signals will unfortunately be absent if the particles are neutral. Note though that according to the arguments in section~\ref{sec-QN}, for many representations ($\vv{3}$, $\vv{6}$, $\vv{15}$, etc.) neutral particles would be exactly stable and are likely already excluded.

In this section we will also present approximate current limits based on recent LHC searches. For simplicity, since the signal efficiencies (including acceptance) are $\cO(1)$, and since the limits are still changing rapidly as the analyses are being updated with more data, we will only display one representative limit curve per search, without simulating the signal efficiency relevant to each of our cases separately. We will present the cross sections as a function of half the bound state mass, $M/2$, which is roughly the mass $m$ of the constituent particle.\footnote{In the dijet channel where the contribution comes from several bound states with different binding energies, we took $M$ to correspond to the lightest one.  However, all the states are included in the computation of the cross section, each with its own mass, since they will usually contribute to the same peak within the experimental resolution.}

\subsection{$\gamma\gamma$ channel}
\label{sec-gmgm}

Any particle with charge and color, no matter what its spin, can be produced in pairs (in $gg$ collisions) in an S-wave $J=0$ color-singlet bound state that can then decay as a typically narrow $\gamma \gamma$ resonance. For spin-$1$ particles, S-wave $J=2$ color-singlet bound states contribute as well.  Even though the dominant decay in all cases is back to a $gg$ pair, the $\gamma\gamma$ signal benefits from smaller backgrounds and better resolution.

\begin{figure}[t]
\begin{center}
\includegraphics[width=0.48\textwidth]{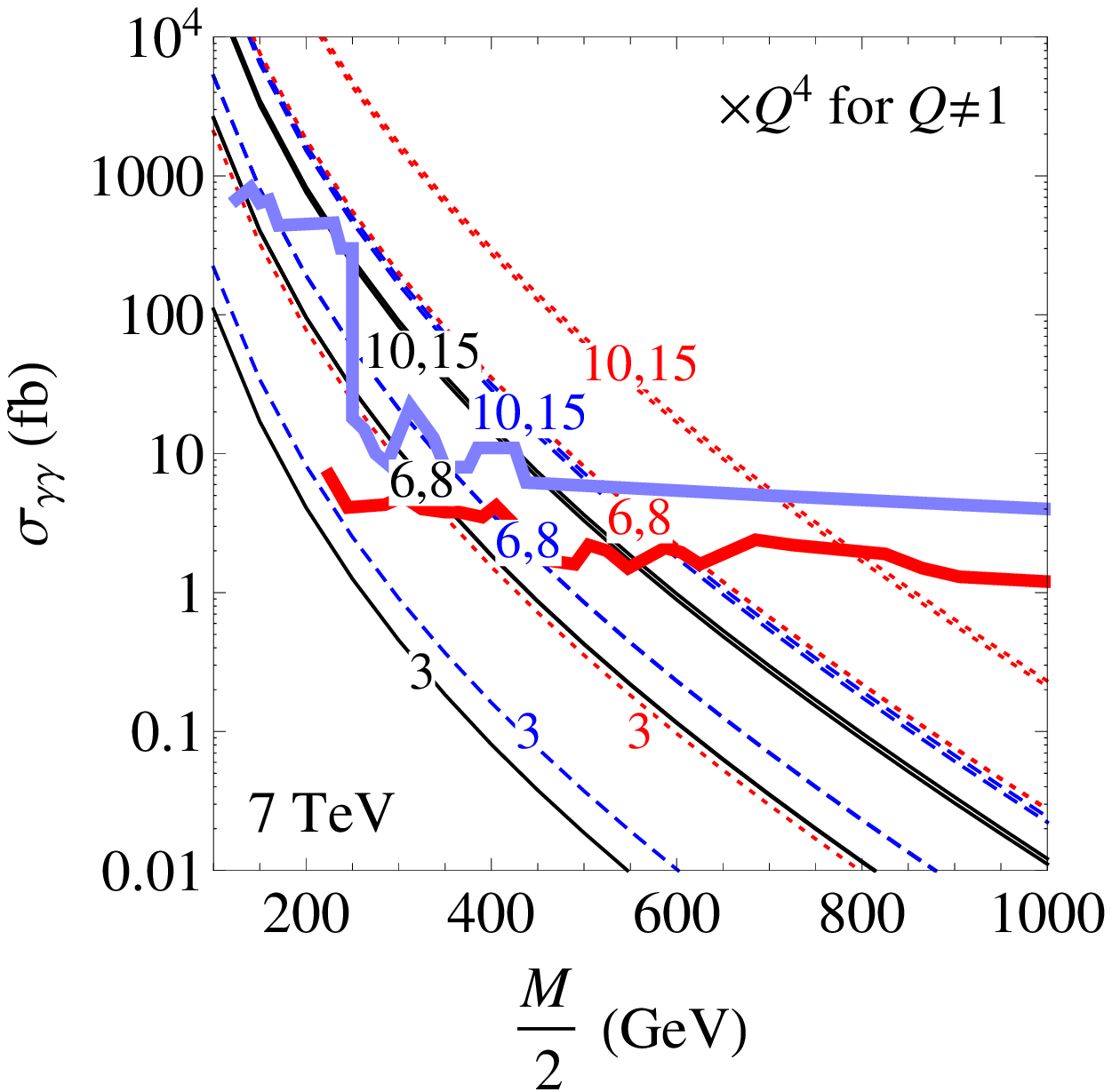}\q
\includegraphics[width=0.49\textwidth]{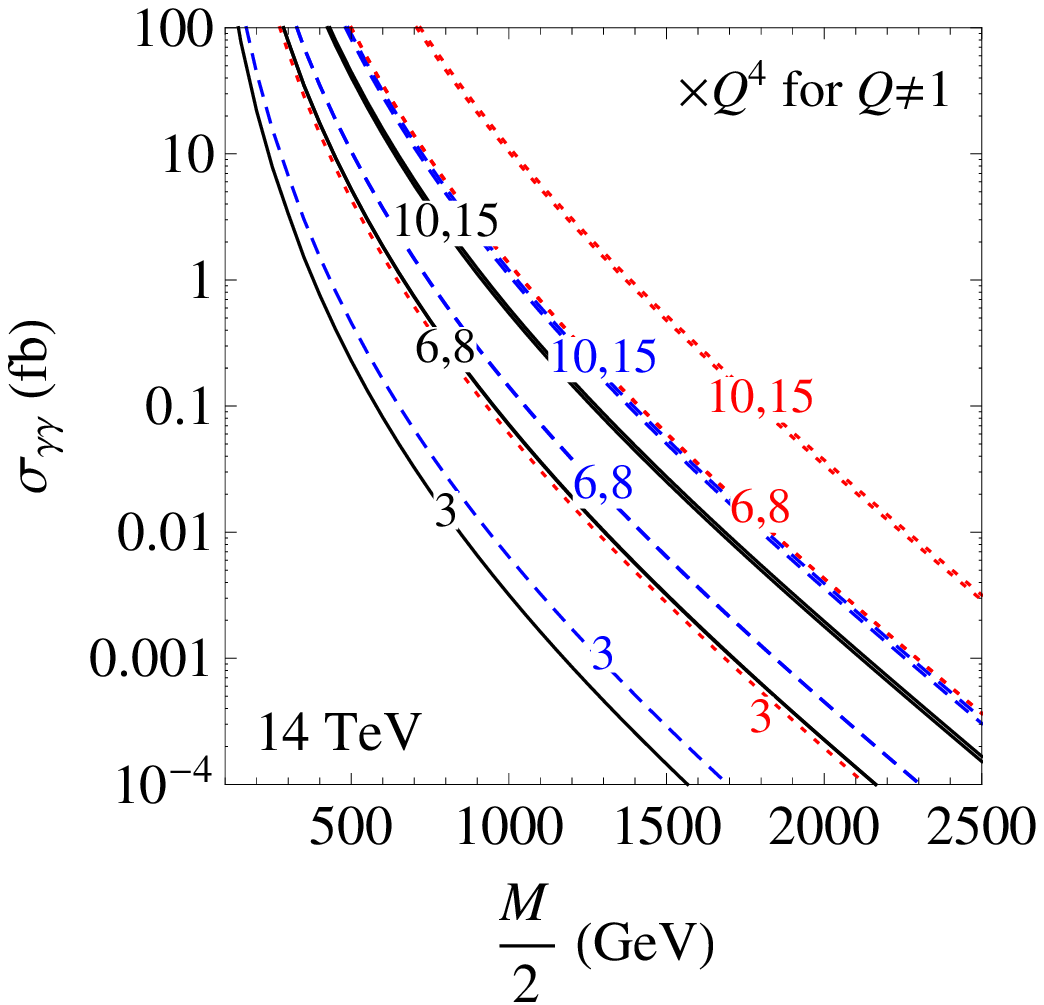}
\caption{$\gamma\gamma$ signal cross sections at the 7~TeV (left) and 14~TeV (right) LHC for bound states of particles with spin $j=0$ (solid black), $\frac12$ (dashed blue) or $1$ (dotted red) and charge $Q = 1$. For other values of $Q$ they need to be multiplied by $Q^4$. Our estimates of the current $95\%$~CL exclusion limits are the thick red curve (ATLAS, 4.9~fb$^{-1}$~\cite{ATLAS-CONF-2012-087}), and the thick blue curve (CMS, 1.14~fb$^{-1}$~\cite{CMS-PAS-EXO-11-038}) which is supplemented for $M/2 < 250$~GeV by a limit from a lower luminosity search (CMS, $36$~pb$^{-1}$~\cite{CMS-PAS-EXO-10-019}).}
\label{fig-cs-diphoton}
\end{center}
\end{figure}

The cross section of the $\gamma\gamma$ signal due to a spin-$J$ bound state is\footnote{The $\gamma\gamma$ annihilation signal has been computed in the past for bound states of color triplets (see, e.g.,~\cite{Martin:2008sv,Martin:2009dj,Younkin:2009zn,Kahawala:2011pc} for scalars,~\cite{Pancheri:1992km,Barger:1987xg,Kahawala:2011pc} for fermions, and~\cite{Kahawala:2011pc} for vectors) and color octets (see, e.g.,~\cite{Kim:2008bx} for scalars).}
\be\label{sigma-gamma-gamma0}
\sigma_{\gamma\gamma} \simeq \sigma_{gg\to\BS}\,\frac{\Gamma_{\BS\to\gamma\gamma}}{\Gamma_{\BS\to gg}}
= \frac{\l(2J+1\r)C_R^3}{2^{14}}\,
\bar\alpha_s^3\, \frac{\cL_{gg}(M^2)}{M^2}\int_0^\pi d\theta\,\sin\theta\,\sum_{\varepsilon_1,\varepsilon_2}\l|\cM_{X\bar X \leftrightarrow\gamma\gamma}(\theta)\r|^2
\ee
Here ${\cal M}_{X\bar X\leftrightarrow\gamma\gamma}$ is the matrix element for free $X$ and $\bar X$ to convert to a pair of photons and the sum is over the photon polarizations.  Also appearing in this expression is the parton luminosity for a pair of partons $a$ and $b$, defined as
\be
\cL_{ab}(\hat s) = \frac{\hat s}{s}\,\int_{\hat s/s}^1 \frac{dx}{x}\, f_{a/p}(x)\, f_{b/p}\l(\frac{\hat s}{xs}\r) \ ,
\ee
where $\sqrt s$ is the collider center-of-mass energy. The resulting $\gamma\gamma$ signal from the bound states is
\be\label{sigma-gamma-gamma}
\sigma_{\gamma\gamma} = \frac{Q^4 C_R^3 D_R}{64}\,
\pi^2\alpha^2 \bar\alpha_s^3\, \frac{\cL_{gg}(M^2)}{M^2}
\times \l\{1,\, 2,\, 19\r\}\qq\mbox{for $j=\l\{ 0, \frac12, 1 \r\}$}
\ee
We have evaluated\footnote{Here and in the following, we are using the NLO MSTW 2008 PDFs~\cite{Martin:2009iq} evaluated at the scale $m$.} this for the LHC in figure~\ref{fig-cs-diphoton}, which also shows our estimate of the exclusion limit from ATLAS~\cite{ATLAS-CONF-2012-087} and CMS~\cite{CMS-PAS-EXO-10-019,CMS-PAS-EXO-11-038}.\footnote{Unfortunately the high-luminosity ATLAS and CMS searches~\cite{ATLAS-CONF-2012-087,CMS-PAS-EXO-11-038} have not presented limits for $M < 450$~GeV or $M < 500$~GeV, respectively (even though good data seems to be available down to at least $M \approx 200$~GeV in both cases; and of course Higgs boson searches work at even lower masses). As a result, we could only present limits from the much earlier 36~pb$^{-1}$ CMS search~\cite{CMS-PAS-EXO-10-019} in that range of masses. We also could not make use of the CMS search with 2~fb$^{-1}$~\cite{Chatrchyan:2011fq} since it has not presented limits on the cross section times branching ratio.}

For $j=0$ or $\frac12$, the signal is coming from $J=0$ bound states and is therefore isotropic. For $j=1$, both $J=0$ and $J=2$ bound states contribute. The signal from $J=2$ bound states, which contributes $16/19$ of the cross section, will have the angular distribution
\be
\frac{1}{\Gamma_{\BS\to\gamma\gamma}}\,\frac{d\Gamma_{\BS\to\gamma\gamma}}{\sin\theta\,d\theta} = \frac{5}{4}\l(\cos^8\frac\theta2 + \sin^8\frac\theta2\r)
\label{angular-J2}
\ee
where $\theta$ specifies the direction of motion of the photons relative to the beam axis in the center of mass frame. This angular dependence should allow us to distinguish between bound states of spin-$1$ particles and those of spin-$0$ or spin-$1/2$ particles.

\subsection{$\gamma+$jet channel}

For particles in color representations higher than the triplet, the potential is attractive also in the color-octet state.  Color-octet bound states can decay to a photon and a gluon, producing a photon-jet resonance.  Perhaps surprisingly, we find that these states can often be observable.  For spin-0 and spin-1 constituents, these states are especially important, as di-lepton resonances are rarely produced, as we will see.  Happily, measuring the rates for the two-photon and photon-jet resonances provides complementary information about the charge $Q$ and the representation $R$.

An octet state can be constructed from a pair of particles in any representation $R$ by using the color wavefunction
\be
\frac{1}{\sqrt{T_R}} \l(T_R^a\r)^j_i
\label{octet-wavefunction}
\ee
where $\l(T_R^a\r)^j_i$ are the generators and $T_R$ in the prefactor is the index of the representation. For some representations there is more than one way to form an octet ({\it e.g.}, for  $R= \vv{8}$, one can use $d^{abc}$ as well as $f^{abc} \propto \l(T_R^a\r)^{bc}$). However, the $\gamma g$ matrix element turns out to be proportional to $\l(T_R^a\r)^j_i$, so color octets formed with wavefunctions other than (\ref{octet-wavefunction}) do not contribute to the $\gamma+$jet signal.

\begin{figure}[t]
\begin{center}
\includegraphics[width=0.47\textwidth]{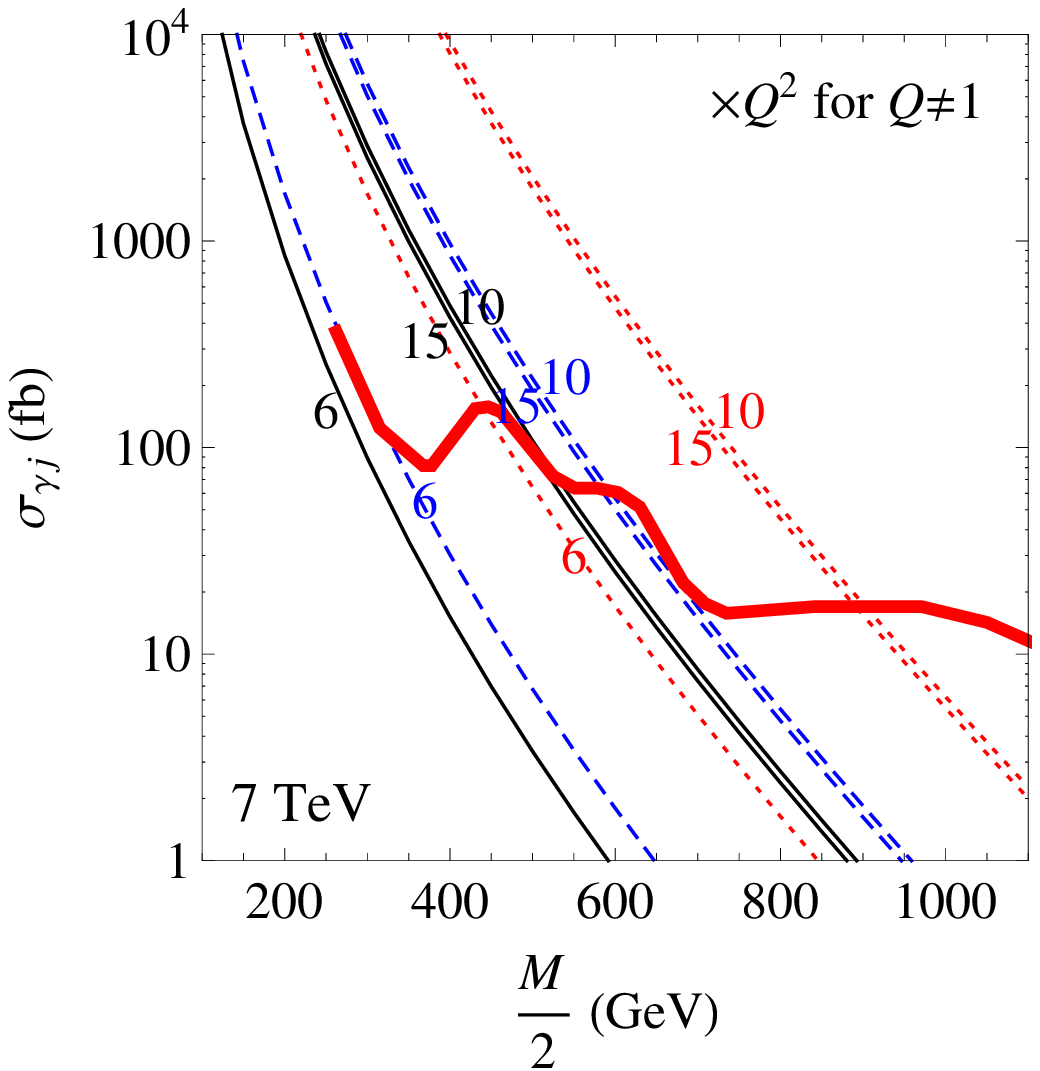}\q
\includegraphics[width=0.49\textwidth]{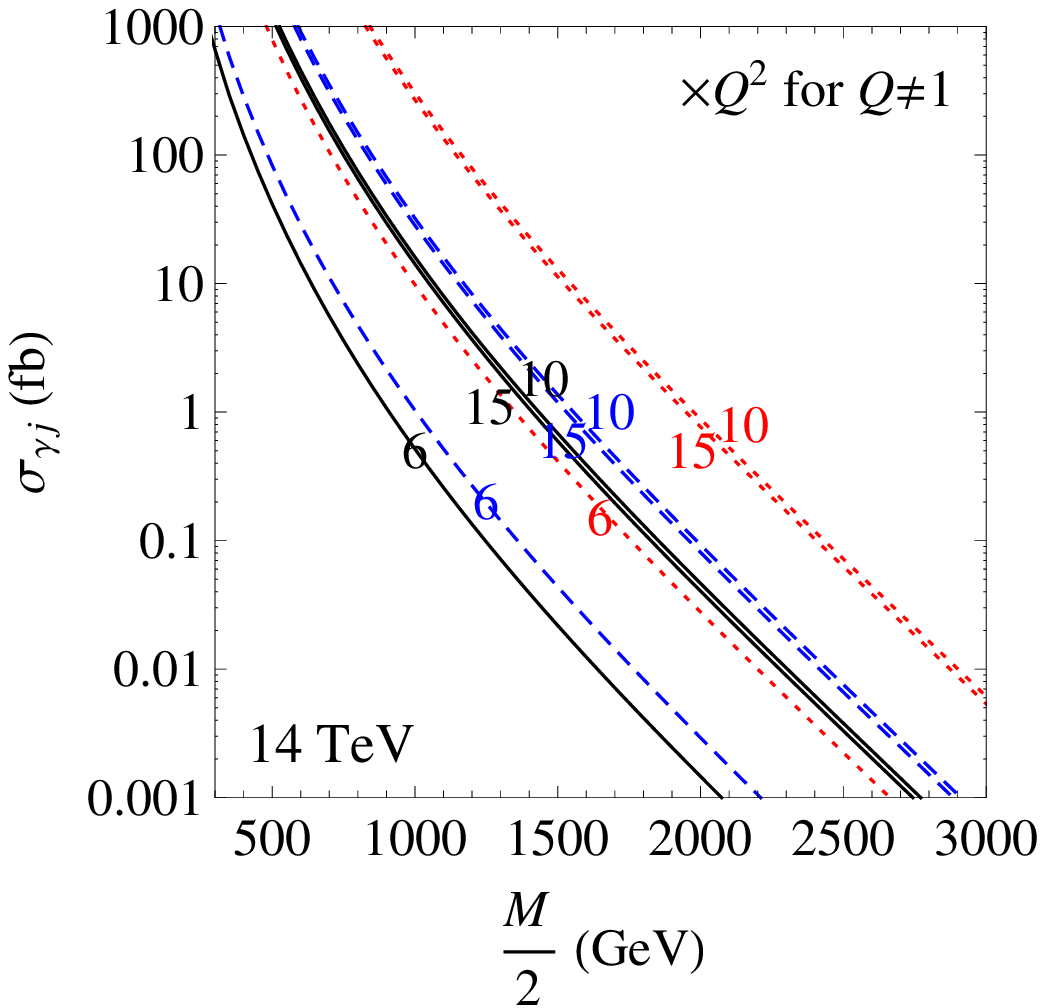}
\caption{$\gamma+$jet signal cross sections at the 7~TeV (left) and 14~TeV (right) LHC for bound states of particles with spin $j=0$ (solid black), $\frac12$ (dashed blue) or $1$ (dotted red) and charge $Q = 1$. For other values of $Q$ they need to be multiplied by $Q^2$. The thick red curve is our estimate of the $95\%$ CL exclusion limit from ATLAS~\cite{Collaboration:2011tg} (with 2.11~fb$^{-1}$).}
\label{fig-cs-photon-jet}
\end{center}
\end{figure}

The matrix elements for production from $gg$ (for particles of any spin $j$) are proportional to $\l\{T_R^b,T_R^c\r\}_j^i$ which will be combined in a color trace with (\ref{octet-wavefunction}), so such color-octet states will be produced with a rate proportional (at leading order) to the square of the anomaly coefficient $A_R$. Unless $A_R$ vanishes (as happens for $R = \vv{8}$), the results from the previous section, for singlet states annihilating to two photons, carry over directly to this section, with very small adjustments.  Our result (\ref{sigma-gamma-gamma}) is simply modified by the replacements
\be  \ C_R^3 \to 8 \l(C_R - \frac32\r)^3 \  ,\qq
Q^4\, D_R\, \alpha^2 \to Q^2\, T_R\,  \alpha\, \alpha_s \ ,
\ee
The first step in the replacement stems from the change in the potential, which in the singlet state is proportional to $C_R$ and in the octet state to $C_R-\frac 32$; see eq.~(\ref{potential}).  This propagates, cubed, into the squared wave function at the origin, $|\psi(\vv{0})|^2$.  Finally there is a factor of 8 due to the octet multiplicity. The second part of the replacement rule arises from the differences in the matrix elements (contracted with the color wavefunctions), which for $\gamma\gamma$ are proportional to
\be
e^2 Q^2\, \delta^i_j \cdot \frac{\delta^j_i}{\sqrt{D_R}} = e^2 Q^2 \sqrt{D_R}
\ee
and for $\gamma g$ to
\be
e Q\, g_s \l(T_R^b\r)^i_j \cdot \frac{1}{\sqrt{T_R}}\l(T_R^a\r)^j_i = eQ\, g_s \sqrt{T_R}\,\delta^{ab}
\ee
We thus get
\be
\sigma_{\gamma g} = \frac{Q^2 \l(C_R-\frac{3}{2}\r)^3 T_R}{8}\,
\pi^2 \alpha\, \alpha_s\, \bar\alpha_s^3\, \frac{\cL_{gg}(M^2)}{M^2} \times \l\{1,\, 2,\, 19\r\}\qq\mbox{for $j=\l\{ 0, \frac12, 1 \r\}$}
\label{sigma-gamma-g}
\ee
As before, the factor of 19 involves 3 from the $J=0$ state and 16 from the $J=2$ state.  The resulting LHC cross sections are shown in figure~\ref{fig-cs-photon-jet}, which also shows our estimate of the exclusion limit from ATLAS~\cite{Collaboration:2011tg}.\footnote{No limits on narrow $\gamma g$ resonances have been presented as yet.  Such resonances can be simulated by using \textsc{FeynRules}~\cite{Christensen:2008py} to create a UFO model~\cite{Degrande:2011ua} which can be provided as an input to \textsc{MadGraph~5}~\cite{Alwall:2011uj}. To simulate the search~\cite{Collaboration:2011tg}, we showered the events in \textsc{Pythia~8}~\cite{Sjostrand:2007gs}, applied the photon isolation criterion, jet algorithm, photon reconstruction efficiency of $85\%$, and all the cuts. To obtain the exclusion curve, we used the limit of~\cite{Collaboration:2011tg} on generic Gaussian-shape peaks following the procedure explained in the appendix of the analogous dijet search~\cite{Aad:2011fq}, including cutting the non-Gaussian tail, adjusting the mass, and determining the width of the effective Gaussian. We found the acceptance (times efficiency) to be $\approx 33\%$ and the width (dominated by showering, but including also the detector resolution from~\cite{Collaboration:2011tg}) to vary between 8\% at low masses and 7\% at high masses. We used the limit from~\cite{Collaboration:2011tg} which assumes 7\% width.} The angular distributions of the annihilation products are the same as in the $\gamma\gamma$ channel.

\subsection{Dijet channel}

For particles with any spin and in any color representation, there are S-wave bound states (with $J=0$ and, for $j=1$, also $J=2$) produced via $gg \to \BS$ and annihilating mostly back to $gg$. These include the color-singlet and color-octet bound states discussed in the previous two subsections, as well as bound states in the $\vv{27}$. The result is a resonant dijet signal. For $j=\frac12$ there is also a comparable contribution from S-wave $J=1$ color-octet bound states produced via $q\bar q \to \BS$ and annihilating to $q\bar q$. The annihilation produces all quark flavors equally (unless limited by phase space), so $1/6$ of the cases are actually $t\bar t$ rather than ordinary jets.

The squared matrix element for pair production of $X\bar X$ from $gg$ at the threshold is proportional to
\be
|\cM|^2 \propto \mbox{Tr}\l(\l\{T_R^a, T_R^b\r\} \l\{T_R^c, T_R^d\r\}\r)
\ee
where $a,b$ and $c,d$ are the color indices of the gluons from $\cM$ and $\cM^\ast$, respectively, and as usual $R$ is the representation of $X$.  The $X\bar X$ bound state in a representation ${\cal R}$ can only be produced when the gluon pairs are each in that representation.  We may project the gluon pairs onto the representation ${\cal R}$ using projection operators given in eq.~(B.15) of~\cite{Beneke:2009rj}. Direct calculation yields
\bea
P_\vv{1}^{abcd}\,\mbox{Tr}\l(\l\{T_R^a, T_R^b\r\} \l\{T_R^c, T_R^d\r\}\r)
&=& \frac12C_R^2 D_R \\
P_\vv{8_S}^{abcd}\mbox{Tr}\l(\l\{T_R^a, T_R^b\r\} \l\{T_R^c, T_R^d\r\}\r)
&=& \frac45\l(C_R + \frac34\r)C_R D_R \\
P_\vv{27}^{abcd}\,\mbox{Tr}\l(\l\{T_R^a, T_R^b\r\} \l\{T_R^c, T_R^d\r\}\r)
&=& \frac{27}{10}\l(C_R - \frac43\r)C_R D_R
\eea
with projection onto $\vv{8_A}, \vv{10}, \vv{\bar{10}}$ giving zero.

\begin{figure}[t]
\begin{center}
\includegraphics[width=0.47\textwidth]{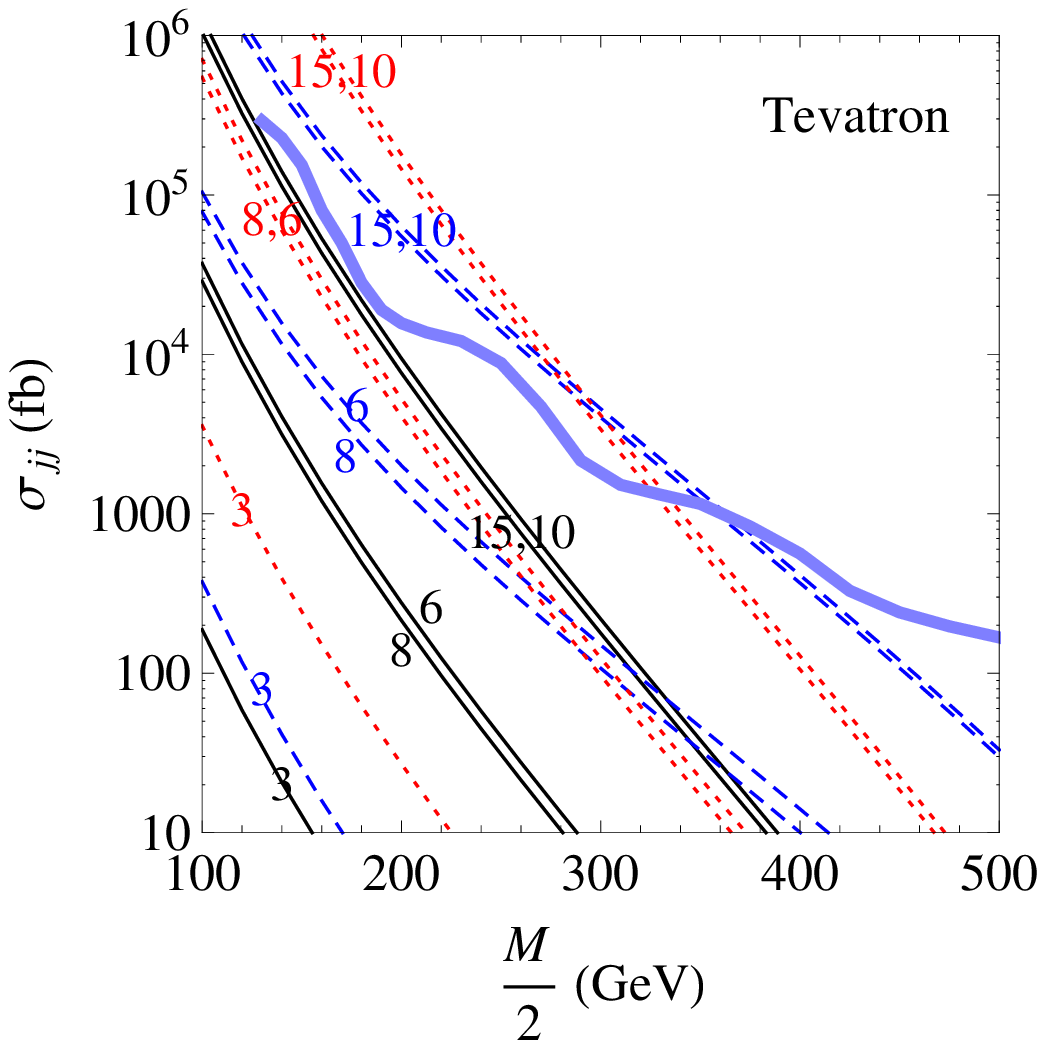}\\\vspace{3mm}
\includegraphics[width=0.46\textwidth]{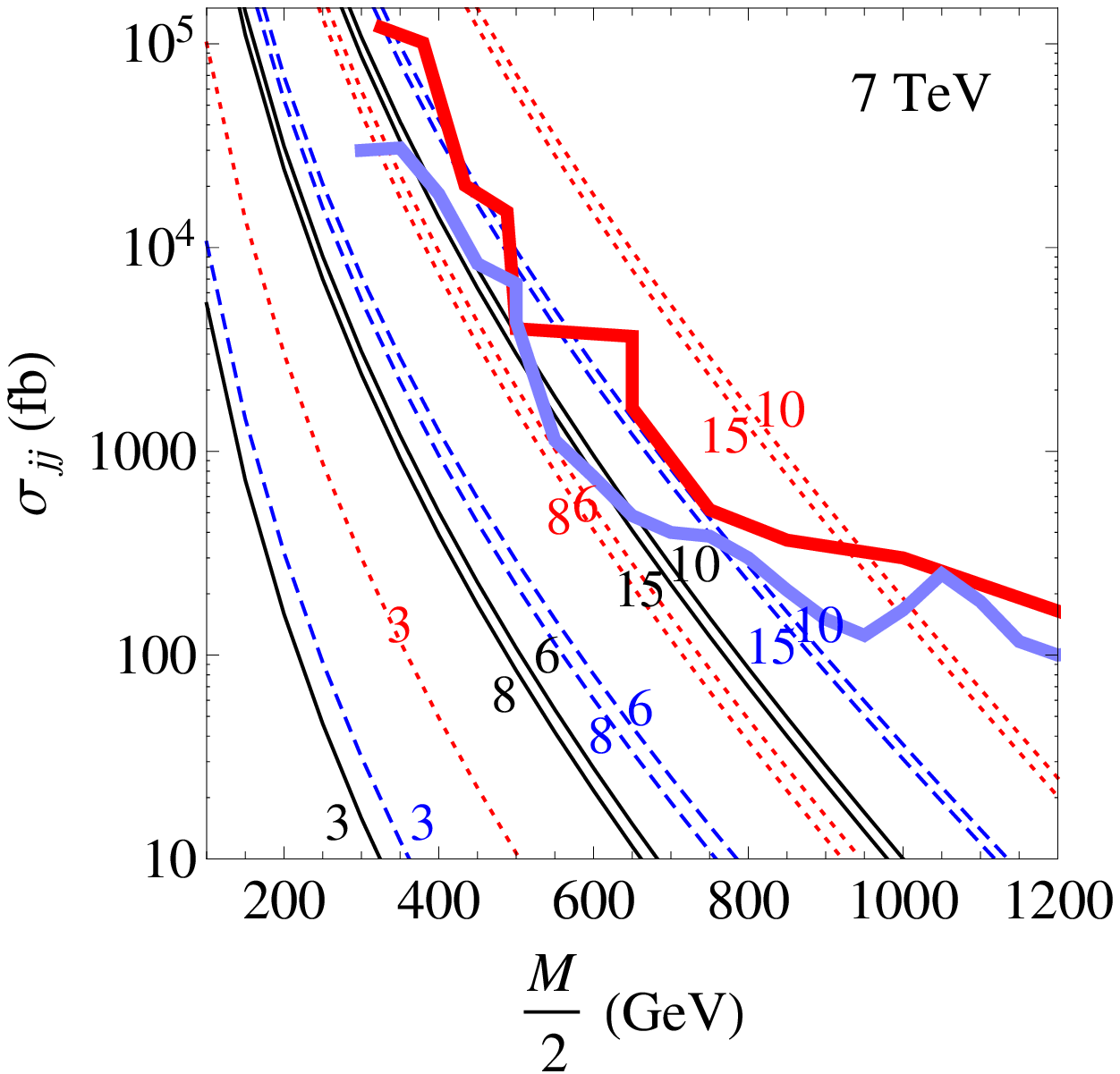}\q
\includegraphics[width=0.47\textwidth]{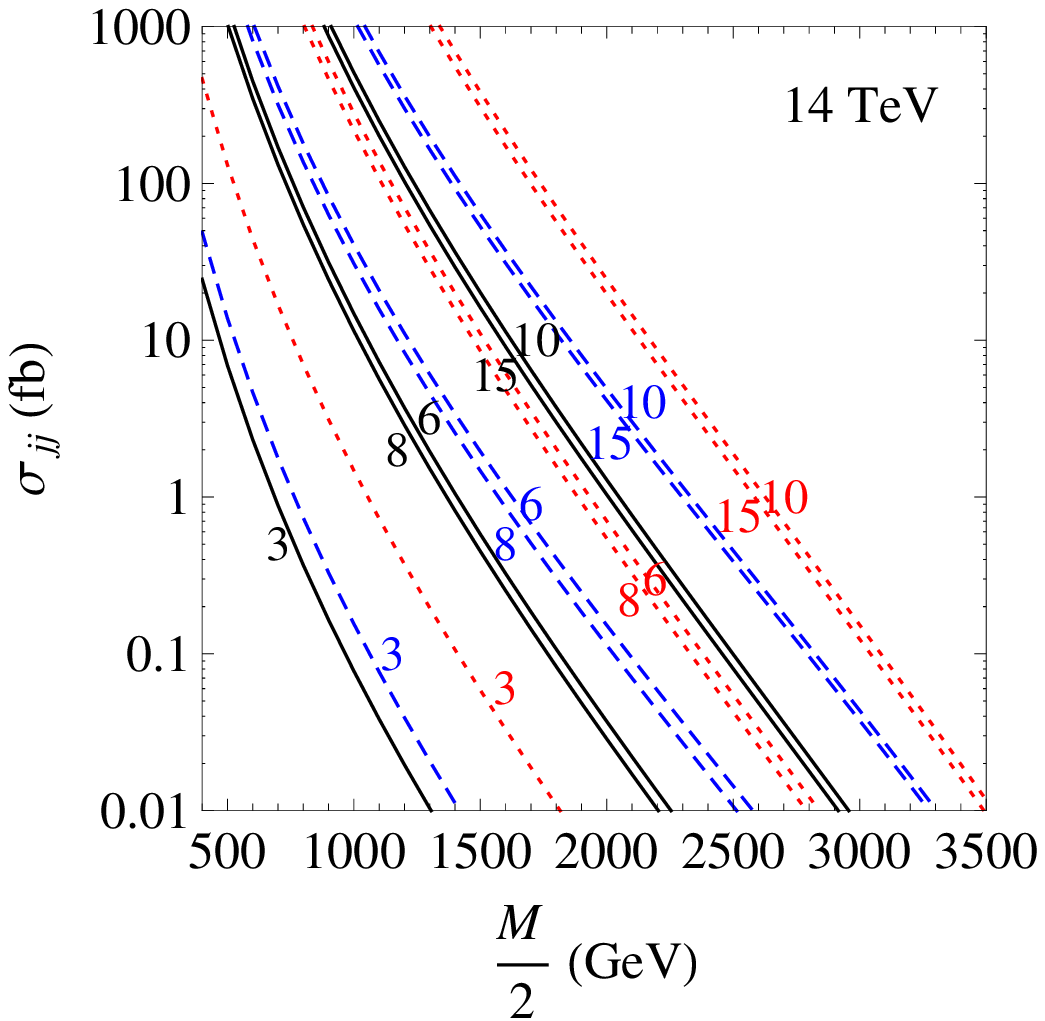}
\caption{Dijet signal cross section at the Tevatron (top), 7~TeV LHC (bottom left) and 14~TeV LHC (bottom right) for bound states of particles with spin $j=0$ (solid black), $\frac12$ (dashed blue) or $1$ (dotted red). For the Tevatron, the thick blue curve is the limit from CDF~\cite{Aaltonen:2008dn} (with 1.13~fb$^{-1}$). For the 7~TeV LHC, the thick red curve is the limit from ATLAS~\cite{ATLAS-CONF-2012-038} (with 4.8~fb$^{-1}$), supplemented for $M/2 < 650$~GeV by limits from earlier ATLAS searches with lower luminosity~\cite{Aad:2011fq,Aad:2011aj} (with 1~fb$^{-1}$ and 36~pb$^{-1}$, respectively), and the thick blue curve is the limit from CMS~\cite{CMS-PAS-EXO-11-094} (with 5~fb$^{-1}$ for $M/2 > 500$~GeV, and 0.13~fb$^{-1}$ analyzed with a special technique for lower masses).}
\label{fig-cs-dijet}
\end{center}
\end{figure}

From this and table~\ref{tab-reps} we see that there will be color-singlet bound states for particles in any representation $R$, color-octet bound states for all representations except for $\vv{3}$, and color-$\vv{27}$ bound states for all $R$ except for $\vv{3}$, $\vv{8}$ and $\vv{6}$. Their production cross sections are all proportional to the easily-obtained production rate for the color singlet, and thus we have\footnote{Previous results for the $gg$ channel covered bound states of color triplets (see, e.g.,~\cite{Martin:2008sv,Martin:2009dj,Younkin:2009zn,Kahawala:2011pc} for $j=0$ and~\cite{Pancheri:1992km,Barger:1987xg,Kahawala:2011pc} for $j=\frac12$) and color octets (see, e.g.,~\cite{Kim:2008bx} for $j=0$).}
\be
\sigma^{gg}_{jj,\vv{1}} = \frac{D_R C_R^5}{512}\,
\pi^2\alpha_s^2 \bar\alpha_s^3\, \frac{\cL_{gg}(M^2)}{M^2}
\times \l\{1,\, 2,\, 19\r\}\q\mbox{for $j=\l\{ 0, \frac12, 1 \r\}$}
\label{xsec-gg-dijet-1}
\ee
\be
\sigma^{gg}_{jj,\vv{8}} = \frac{D_R C_R\l(C_R + \frac34\r)\l(C_R - \frac32\r)^3}{320}\,
\pi^2\alpha_s^2 \bar\alpha_s^3\, \frac{\cL_{gg}(M^2)}{M^2}
\times \l\{1,\, 2,\, 19\r\}\q\mbox{for $j=\l\{ 0, \frac12, 1 \r\}$}
\ee
\be
\sigma^{gg}_{jj,\vv{27}} = \frac{27 D_R C_R\l(C_R - \frac43\r)\l(C_R - 4\r)^3}{2560}\,
\pi^2\alpha_s^2 \bar\alpha_s^3\, \frac{\cL_{gg}(M^2)}{M^2}
\times \l\{1,\, 2,\, 19\r\}\q\mbox{for $j=\l\{ 0, \frac12, 1 \r\}$}
\ee
For $j=\frac12$ and any $R$ except for $\vv{3}$, there is an additional contribution to the cross section from the $q\bar q$ channel:
\be
\sigma^{q\bar q}_{jj,\vv{8}} = \frac{D_R C_R\l(C_R - \frac32\r)^3}{9}\,
\pi^2\alpha_s^2 \bar\alpha_s^3\, \frac{\sum_q\cL_{q\bar q}(M^2)}{M^2}
\label{xsec-qqbar-dijet}
\ee
The total dijet cross sections (excluding the $t\bar t$ contribution) for the 7 and 14~TeV LHC (as well as Tevatron) are shown in figure~\ref{fig-cs-dijet}, which also shows our estimates of the exclusion limits from CDF~\cite{Aaltonen:2008dn}, ATLAS~\cite{ATLAS-CONF-2012-038,Aad:2011fq,Aad:2011aj} and CMS~\cite{CMS-PAS-EXO-11-094}.\footnote{For the Tevatron, we used the RS graviton limit of~\cite{Aaltonen:2008dn}. For the 7~TeV ATLAS data we used the limit of~\cite{ATLAS-CONF-2012-038} on a color-octet scalar resonance decaying to gluons. Since the mass range of this 4.8~fb$^{-1}$ search is limited from below by the triggers, we used the analogous limit from the earlier 1~fb$^{-1}$ search~\cite{Aad:2011fq} for $M/2 < 650$~GeV, and the yet earlier 36~pb$^{-1}$ search~\cite{Aad:2011aj} for $M/2 < 500$~GeV. In~\cite{Aad:2011aj}, a limit on a color-octet scalar was not available and we used the limit on signals which appear as Gaussians, following the instructions from the appendix of~\cite{Aad:2011fq}. For CMS~\cite{CMS-PAS-EXO-11-094}, we used the limit on a resonance decaying to $gg$ (for $M/2 < 500$~GeV, the data were analyzed with a special approach adapted to the high event rate at low masses, using however only the last 0.13~fb$^{-1}$). For this CMS search, we used their quoted acceptance of ${\cal A} \approx 60\%$ (for isotropic decays), while for the other searches we simulated a scalar $gg$ resonance in \textsc{Pythia}. For~\cite{ATLAS-CONF-2012-038} and~\cite{Aad:2011fq} we get ${\cal A} \approx 60\%$. For~\cite{Aad:2011aj}, ${\cal A} \approx 30\%$, the mean mass of the effective Gaussian is shifted down by $\approx 8.5\%$ relative to the real mass (in our figure, $M$ is the real mass), and its width (dominated by showering, but including detector resolution as well) is $\approx 9\%$ (so we used the limit of~\cite{Aad:2011aj} on 10\%-width resonances). For~\cite{Aaltonen:2008dn}, ${\cal A}$ varies between 50\% at low masses and 75\% at high masses.} It would be beneficial to keep improving the LHC reach in the low-mass region (which is constrained very weakly by the Tevatron), which would involve working in a regime where the trigger is not fully efficient, using a prescaled trigger and/or storing reduced event content as has been done by CMS in~\cite{CMS-PAS-EXO-11-094}.

Note that unlike all the other signals studied in this paper, the dijet signal is present even if the particles are not charged. While we were assuming the fields $X$ to be complex, if $Q=0$ and the color representation is real (such as $\vv{8}$), the field $X$ may be real. In such case the cross sections (\ref{xsec-gg-dijet-1})--(\ref{xsec-qqbar-dijet}) would be half as big.\footnote{The resulting expressions would apply, for example, to bound states of gluinos, see e.g.~\cite{Cheung:2004ad,Kahawala:2011pc}, or KK gluons~\cite{Kahawala:2011pc}, in the limit that the squarks or the KK quarks are heavy.}

The angular distributions of the annihilation products are the same as in the $\gamma\gamma$ channel, except for $j=\frac12$ particles where there is an additional contribution from the spin-1 bound states of the $q\bar q$ channel with the angular distribution
\be
\frac{1}{\Gamma_{\BS\to q\bar q}}\,\frac{d\Gamma_{\BS\to q\bar q}}{\sin\theta\,d\theta} = \frac{3}{8}\l(1 + \cos^2\theta\r)
\label{angular-J1}
\ee

\subsection{$\ell^+\ell^-$ channel}
\label{sec-dilepton}

Spin-$\frac12$ particles in any color representation can form color-singlet $J=1$ S-wave bound states analogous to the $J/\psi$ and $\Upsilon$. The interesting signal of these bound states is their annihilation into a pair of leptons (via an $s$-channel photon or $Z$). They cannot be produced directly from a pair of gluons or quarks (except by electroweak interactions), so we need to consider the various subdominant production processes. Besides the electroweak production from $q\bar q$, these include production from $gg$ in association with a gluon (or a photon or $Z$) and production via a radiative transition of a color-singlet or color-octet P-wave bound state (along with a soft photon or gluon). The P-wave bound states themselves are produced directly, although with a suppressed rate, from a pair of gluons.

It turns out that any of these processes can be important, depending on the representation, charge and collider energy.  This is illustrated in table~\ref{tab-prod-fractions} for the case of $m = 1$~TeV. The ratios between the processes also vary strongly as a function $m$, primarily because the different processes depend on different PDFs. And of course one must keep in mind that, like everything in this paper, the ratios are calculated at leading order and may change significantly upon including higher order corrections.

The $p_T$ distribution of the bound state is an interesting observable, and it is sensitive to the relative contributions of the different production modes, which will have different $p_T$ spectra.  In particular, the $gg\to{\cal B}g$ process provides a hard gluon against which the bound state can recoil, unlike the electroweak production or the radiative processes where a hard recoiling gluon can come only from initial state radiation.  Consequently, a measurement of the $p_T$ distribution is sensitive to the quantum numbers of the state.  But methods much more sophisticated than ours would be needed to predict this observable.

We will now analyze the production mechanisms one by one, then compute the dilepton branching ratio of the bound state and plot the resulting signal, and at the end comment on the much smaller signals expected from bound states of spin-0 or spin-1 particles.

\begin{table}[t]
$$\begin{array}{|c||c|c|c|c|c||c|c|c|c|c|}\hline
\mbox{7 TeV LHC} & \multicolumn{5}{|c||}{Q = \frac23} & \multicolumn{5}{|c|}{Q = 2} \\\hline
\mbox{$m = 1$ TeV} &\;\,\vv{3}\,\;&\;\,\vv{8}\,\;&\;\,\vv{6}\,\;&\;\vv{15}\;&\;\vv{10}\;
                   &\;\,\vv{3}\,\;&\;\,\vv{8}\,\;&\;\,\vv{6}\,\;&\;\vv{15}\;&\;\vv{10}\;\\\hline
q\bar q \to \BS                          & 97 & 92 & 71 & 68 & 22 & 99 & 96 & 92 & 88 & 66 \\
gg \to \BS g                             &  2 &  - & 14 &  9 & 24 &  0 &  - &  2 &  1 &  8 \\
gg \to \BS (\gamma/Z)                    &  0 &  1 &  1 &  2 &  1 &  0 &  1 &  1 &  3 &  3 \\
gg \to \,^3P_J^{(\vv{1})} \to \BS\gamma  &  2 &  7 &  8 &  5 &  3 &  0 &  3 &  4 &  5 &  6 \\
gg \to \,^3P_J^{(\vv{8})} \to \BS g      & -  &  - &  6 & 16 & 50 &  - &  - &  1 &  2 & 16 \\\hline
\end{array}$$\vskip 5pt
$$\begin{array}{|c||c|c|c|c|c||c|c|c|c|c|}\hline
\mbox{14 TeV LHC} & \multicolumn{5}{|c||}{Q = \frac23} & \multicolumn{5}{|c|}{Q = 2} \\\hline
\mbox{$m = 1$ TeV} &\;\,\vv{3}\,\;&\;\,\vv{8}\,\;&\;\,\vv{6}\,\;&\;\vv{15}\;&\;\vv{10}\;
                   &\;\,\vv{3}\,\;&\;\,\vv{8}\,\;&\;\,\vv{6}\,\;&\;\vv{15}\;&\;\vv{10}\;\\\hline
q\bar q \to \BS                          & 87 & 77 & 36 & 35 &  7 & 97 & 87 & 73 & 65 & 33 \\
gg \to \BS g                             &  8 &  - & 40 & 24 & 40 &  1 &  - &  9 &  5 & 22 \\
gg \to \BS (\gamma/Z)                    &  1 &  5 &  3 &  7 &  2 &  1 &  5 &  6 & 13 &  8 \\
gg \to \,^3P_J^{(\vv{1})} \to \BS\gamma  &  5 & 18 & 12 &  8 &  2 &  1 &  8 & 10 & 12 & 10 \\
gg \to \,^3P_J^{(\vv{8})} \to \BS g      &  - &  - & 10 & 26 & 49 &  - &  - &  2 &  6 & 27 \\\hline
\end{array}$$
\caption{The fraction (in $\%$) contributed by each of the production mechanisms for color-singlet $J=1$ S-wave bound states ($\BS$) of spin-$\frac12$ particles at the 7~TeV LHC (top table) and 14~TeV LHC (bottom table). The numbers are presented for bound states of $SU(2)$ singlets with $Q = \frac23$ (left) or $2$ (right) and $m = 1$~TeV. Note these numbers change significantly as a function of $m$ as well as $R$ and $Q$.}
\label{tab-prod-fractions}
\end{table}

\subsubsection{Electroweak production from $q\bar q$}

Given that we mostly consider $X\bar X$ states with mass far above the $Z$, we can largely ignore the $Z$ mass to our level of approximation, and write
the electroweak production cross section via an $s$-channel photon or $Z$ as
\be
\sigma = \frac{\pi^2}{108}\, D_R C_R^3 Q^2\,
\frac{\alpha^2 \bar\alpha_s^3}{\cos^4\theta_W} \l(17\sum_{q=u,c} + 5\sum_{q=d,s,b}\r)\frac{\cL_{q\bar q}(M^2)}{M^2}
\ee

\subsubsection{Production in association with a gauge boson}
\label{sec-bleaching}

Generalizing the results for $J/\psi$ and $\Upsilon$ (see, e.g.,~\cite{Barger:1984qg,Drees:1991ig}), the cross section for production in association with a gluon, $gg \to \BS\,g$, is
\be
\sigma = \frac{5\pi}{192\,m^2}\,\frac{A_R^2 C_R^3}{D_R}\,\alpha_s^3\bar\alpha_s^3 \int_0^1 dx_1 \int_0^1 dx_2\; f_{g/p}(x_1)\, f_{g/p}(x_2)\; I\l(\frac{x_1 x_2 s}{M^2}\r)
\ee
where
\be
I(x) = \theta(x-1)\l[\frac{2}{x^2}\l(\frac{x+1}{x-1} - \frac{2x\ln x}{(x-1)^2}\r) + \frac{2(x-1)}{x(x+1)^2} + \frac{4\ln x}{(x+1)^3}\r]
\label{I(x)}
\ee
Production in association with a photon, $gg \to \BS\,\gamma$, is
described by the same expression with the replacement
\be
\frac{A_R^2}{D_R}\,\alpha_s
\to \frac{3}{20}\,D_R C_R^2\, Q^2\alpha
\ee
Production in association with a $Z$ is given by the expression for the photon times $\tan^2\theta_W$.

\subsubsection{Production via color-singlet P-wave states}

Here we consider the possibility of production through the radiative decays of the lowest color-singlet P states, $\,^3P_J^{(\vv{1})}$ (a.k.a. $\chi_J$) with $J = 0, 2$:
\be
gg \to \,^3P_J^{(\vv{1})} \to \BS\,\gamma
\ee
The binding energy of these P waves is $E_b = -C_R^2\,\bar\alpha_s^2 m/16$, which is about 4 times smaller than of the S wave of interest. The rate of the radiative transitions from a P wave to an S wave is (see, e.g.,~\cite{Novikov:1977dq})
\be
\Gamma(\,^3P_J^{(\vv{1})} \to \BS\gamma) = \frac{4}{9}\,Q^2\bar\alpha\, E_\gamma^3\l|\langle R_S|r|R_P\rangle\r|^2
\sim \frac{128}{6561}\,Q^2 C_R^4\, \bar\alpha\,\bar\alpha_s^4 m
\label{em-transition}
\ee
where the last expression is very approximate, and presented only to illustrate parametric dependence.
The reason is that different values of $\bar\alpha_s$ arise.  In our plots below we evaluate the photon energy
$E_\gamma = E_{b,P} - E_{b,S}$ using $\bar\alpha_s$ defined self-consistently at the average radii of the P-wave and
S-wave states, respectively   (see footnote~\ref{Bohr-radius}).
Meanwhile, we estimate the transition amplitude
\be
\l|\langle R_S|r|R_P\rangle\r|^2 = \frac{2^{17}}{3^9\, C_R^2\,\bar\alpha_s^2\, m^2}
\label{RS1|r|RP1}
\ee
using the average $\bar\alpha_s$ of the two states.

The radiative transitions must compete with annihilation of the P-wave states to $gg$, which have the rates
\be
\Gamma(\,^3P_0^{(\vv{1})} \to gg) = \frac{9}{8}D_R C_R^2\,\alpha_s^2\frac{|R'_P(0)|^2}{m^4}
= \frac{3}{2048} D_R C_R^7\,\alpha_s^2\bar\alpha_s^5 m
\label{3P0-annih}
\ee
\be
\Gamma(\,^3P_2^{(\vv{1})} \to gg) =  \frac{3}{10}D_R C_R^2\,\alpha_s^2\frac{|R'_P(0)|^2}{m^4}
= \frac{1}{2560} D_R C_R^7\,\alpha_s^2\bar\alpha_s^5 m
\label{3P2-annih}
\ee
We obtained these expressions from those of quarkonia (see, e.g.,~\cite{Novikov:1977dq}) with appropriately generalized color factors. Since the P wavefunction vanishes at the origin, the matrix elements are proportional to its derivative, so the rates are suppressed by two additional powers of $\bar\alpha_s$ compared with the S-wave processes. For $Q \approx 1$, the branching ratios for the radiative transitions of the color-singlet P waves are $\cO(1)$ for $R = \vv{3}$, $\vv{6}$, and $\vv{8}$, but become much smaller for higher representations because the annihilation rates are proportional to more powers of the color factors.

Similarly, the production cross sections of $^3P_J^{(\vv{1})}$ are
\be
\sigma = \frac{D_R C_R^7}{2^{13}}\,\pi^2\alpha_s^2\bar\alpha_s^5\, \frac{\cL_{gg}(M^2)}{M^2} \times \l\{\frac{3}{4}\,,\, 1\r\}\q\mbox{for}\q J=\l\{0,\, 2\r\}
\ee

\subsubsection{Production via color-octet P-wave states}

Finally we would like to consider the process
\be
gg \to \,^3P_J^{(\vv{8})} \to \BS\,g
\ee
The binding energy of these color-octet P waves is $E_b = -(C_R - \frac32)^2\,\bar\alpha_s^2 m/16$, which is more than 4 times smaller than for the color-singlet S wave. We need to derive the rate for the chromoelectric dipole transition of an octet to a singlet with the emission of a soft gluon. We can start with the rate of electric dipole transitions with the emission of a photon (see, e.g.,~\cite{Novikov:1977dq})
\be
\Gamma(i\to f\gamma) = \frac{4}{3}\,Q^2\,\bar\alpha\, E_\gamma^3\l|\langle F|\vv{r}|I\rangle\r|^2
\ee
where $|I\rangle$ and $|F\rangle$ are the spatial wavefunctions of the initial state $i$ and the final state $f$. We then make the replacement
\be
eQ\langle F|\vv{r}|I\rangle \;\to\;
g_s\mbox{Tr}\l(\chi_F\,T^a_R\,\chi_I^a\r) \langle F|\vv{r}|I\rangle
\ee
where $\chi_I^a$ and $\chi_F$ are the color wavefunctions of the initial and final state and there is no summation over the color index of the initial state, $a$. Substituting $\chi_F = \delta_{jk}/\sqrt{D_R}$ (where $j,k$ are indices in the representation $R$) we have
\be
\Gamma(i\to f g) = \frac{4}{3}\,\bar\alpha_s\, E_g^3\, \frac{\l|\mbox{Tr}\l(\chi_I^a\, T_R^a\r)\r|^2}{D_R}\l|\langle F|\vv{r}|I\rangle\r|^2
\ee
For the case of $\chi_I^a \propto T_R^a$ as in (\ref{octet-wavefunction}), we have
\be
\Gamma(i\to f g) = \frac{1}{6}\,C_R\,\bar\alpha_s\, E_g^3\l|\langle F|\vv{r}|I\rangle\r|^2
\ee
while for any other possible color-octet wavefunction the result vanishes when $\chi_I^a$ is contracted with $T_R^a$. Specializing to $\,^3P_J^{(\vv{8})}$ (with $J = 0$ or $2$) we get
\be
\Gamma(\,^3P_J^{(\vv{8})} \to \BS g) = \frac{1}{18}\,C_R\,\bar\alpha_s E_g^3\, \l|\langle R_S^{(\vv{1})}|r|R_P^{(\vv{8})}\rangle\r|^2
\sim \frac{16}{6561}\, C_R^4 \frac{\l(C_R+\frac{3}{2}\r)^3\l(C_R-\frac{3}{2}\r)^5}{\l(C_R-\frac{1}{2}\r)^7}\, \bar\alpha_s^5\, m
\label{chr-transition}
\ee
where again, as in (\ref{em-transition}), the last expression is very approximate.  As in the discussion following (\ref{em-transition}), we evaluate the gluon energy $E_g$ using the values of $\bar\alpha_s$ self-consistently determined at the radii of the P-wave and S-wave states (see footnote~\ref{Bohr-radius}). For the explicit power of $\bar\alpha_s$ in the middle expression in~(\ref{chr-transition}) and for the transition amplitude involving $R_P$ and $R_S$ (the P and S radial wavefunctions)
\be
\l|\langle R_S^{(\vv{1})}|r|R_P^{(\vv{8})}\rangle\r|^2 = \frac{2^{17}\,C_R^3\l(C_R-\frac{3}{2}\r)^5}{3^9 \l(C_R-\frac{1}{2}\r)^{10}\bar\alpha_s^2\, m^2}
\ee
we use the average $\bar\alpha_s$ of the two states.

The annihilation rates of the P waves to $gg$, which compete with their radiative transitions, can be obtained by appropriately modifying the various color factors in (\ref{3P0-annih})--(\ref{3P2-annih}):
\be
\Gamma(\,^3P_0^{(\vv{8})} \to gg) = \frac{5}{512} \frac{A_R^2 \l(C_R - \frac{3}{2}\r)^5}{D_R C_R}\,\alpha_s^2\bar\alpha_s^5 m
\label{3P0-8-annih}
\ee
\be
\Gamma(\,^3P_2^{(\vv{8})} \to gg) = \frac{1}{384} \frac{A_R^2 \l(C_R - \frac{3}{2}\r)^5}{D_R C_R}\,\alpha_s^2\bar\alpha_s^5 m
\label{3P2-8-annih}
\ee
The branching ratios for the radiative transitions of the P waves are typically very close to $1$ for $R = \vv{8}$, $\vv{6}$ and $\vv{15}$, and somewhat smaller for $R = \vv{10}$, where the annihilation rates are comparable to the transition rate.

The production cross sections of $^3P_J^{(\vv{8})}$, based on (\ref{3P0-8-annih})--(\ref{3P2-8-annih}), are
\be
\sigma = \frac{5}{768} \frac{A_R^2 \l(C_R - \frac{3}{2}\r)^5}{D_R C_R}\,\pi^2\alpha_s^2\bar\alpha_s^5\, \frac{\cL_{gg}(M^2)}{M^2} \times \l\{\frac{3}{4}\,,\, 1\r\}\q\mbox{for}\q J=\l\{0,\, 2\r\}
\ee

\subsubsection{Dilepton branching ratio}
\label{sec-spin1-ann-rates}

\begin{figure}[t]
\begin{center}
\includegraphics[width=0.48\textwidth]{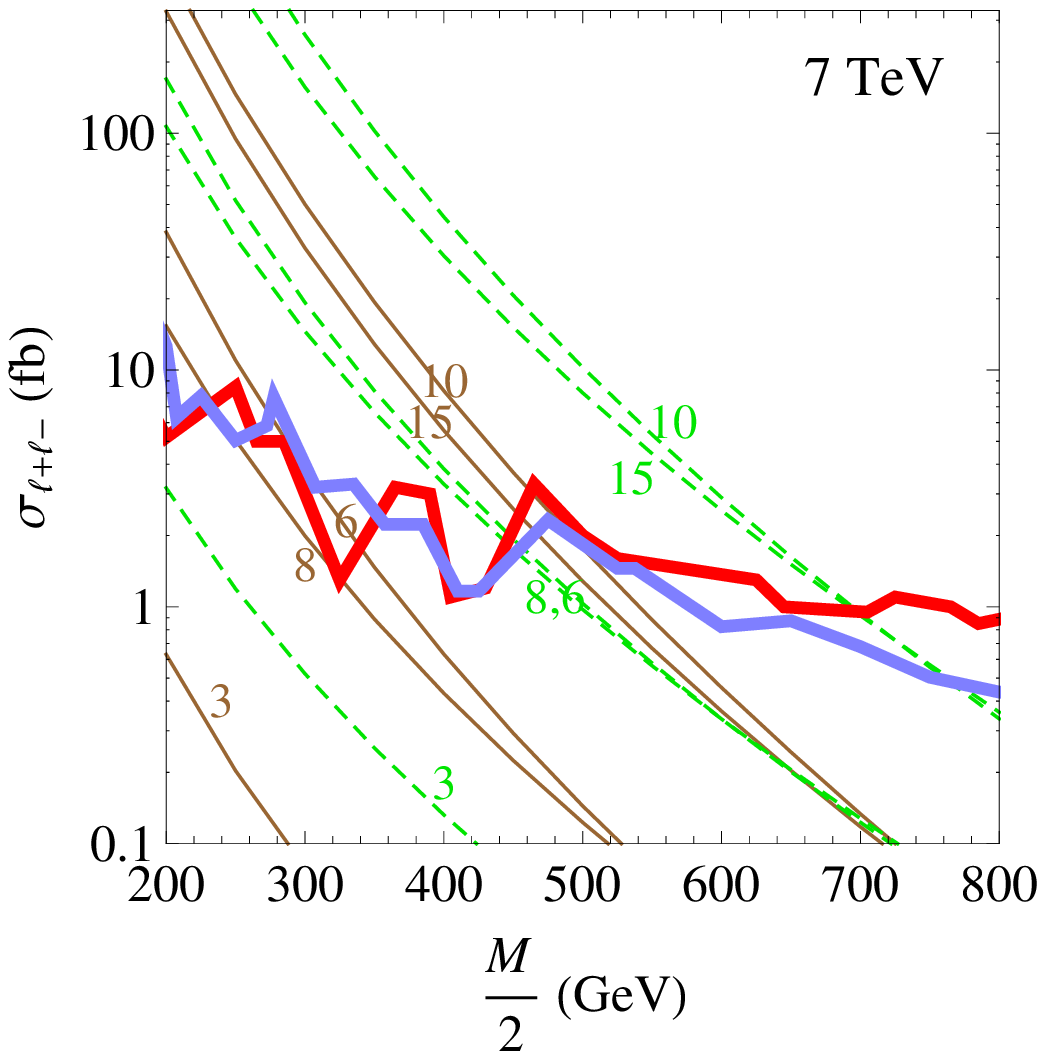}\q
\includegraphics[width=0.485\textwidth]{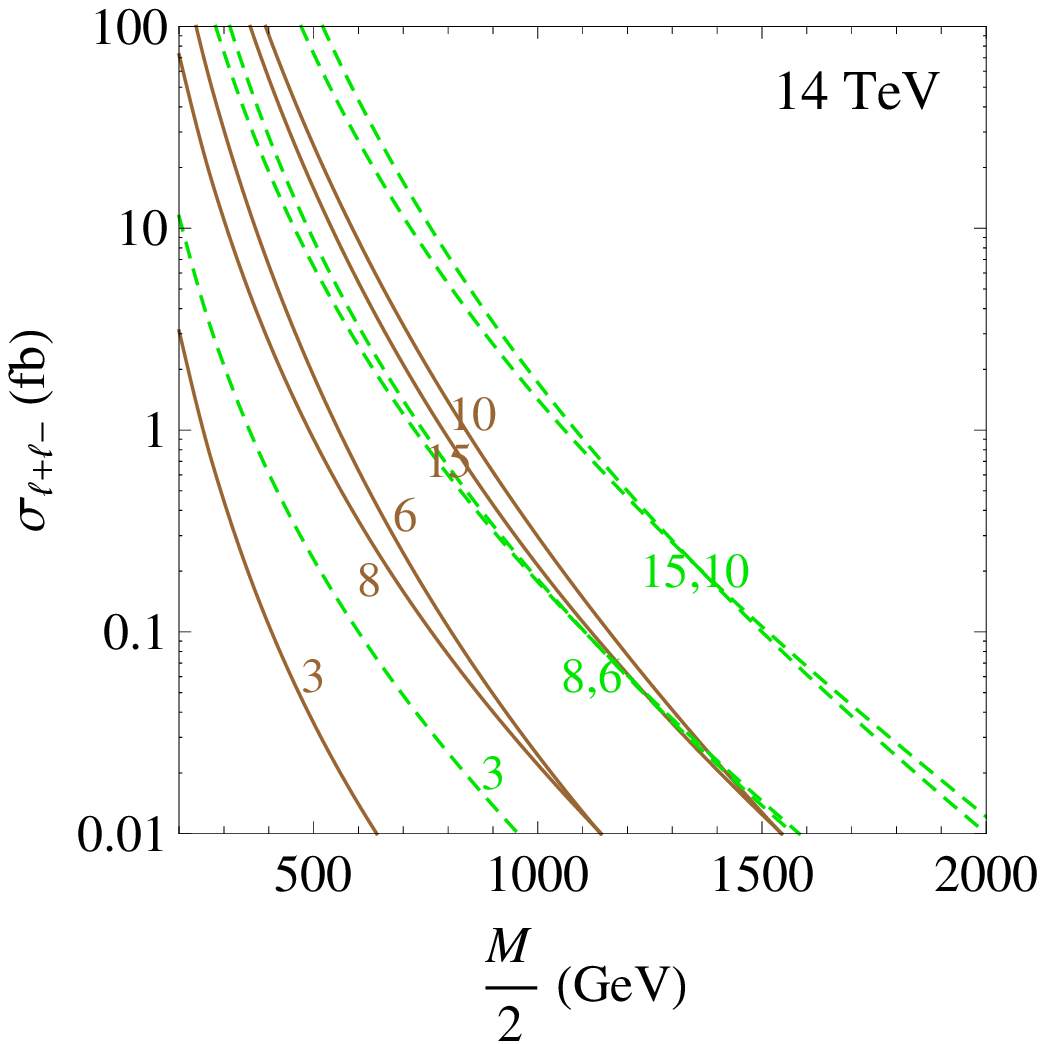}
\caption{$\ell^+\ell^-$ signal cross section (for any single flavor of leptons) at the 7~TeV (left) and 14~TeV (right) LHC for bound states of spin-$1/2$ particles which are $SU(2)$-singlets with $Q=\frac23$ (solid brown) or $Q=2$ (dashed green). Estimates of the 5~fb$^{-1}$, 7~TeV LHC $95\%$ CL exclusion limits are shown in thick red (ATLAS~\cite{ATLAS-CONF-2012-007}) and thick blue (CMS~\cite{Chatrchyan:2012it}).}
\label{fig-cs-dilepton}
\end{center}
\end{figure}

To determine the observable rate for the dilepton resonance, we need to know the branching fraction of the S-wave $J=1$ bound state into Standard Model fermions. For a $J=1$ bound state of fermions $X$ with charge $Q$ and hypercharge $Y$, annihilation can proceed through a photon or $Z$. Neglecting the mass of the $Z$ relative to the bound state mass, we can write the annihilation rate using $SU(2)\times U(1)$ states, giving for any flavor of fermions $f_L, f_R$
\be
\Gamma_{\BS\to f\bar f} = \frac{n_c}{12}\,D_R C_R^3\sum_{\sigma=R,L}\l(\frac{Y_{f_\sigma} Y}{\cos^2\theta_W} + \frac{\l(Q_{f_\sigma}-Y_{f_\sigma}\r)\l(Q-Y\r)}{\sin^2\theta_W}\r)^2
\alpha^2\bar\alpha_s^3 m
\label{annih-ffbar}
\ee
where $n_c = 1$ for leptons and $3$ for quarks. In this paper we restrict ourselves to $SU(2)$ singlets, so $Y = Q$ and only the first term appears.

To compute the branching fractions, we must also compute the partial widths for annihilation to $ggg$, $\gamma gg$, and $Zgg$; we believe there are no other decays with considerable rates.\footnote{The bound state can annihilate to $Z$+Higgs via a diagram with an $s$-channel $Z$. However, the rate is small. For example, for the Standard Model Higgs, in the limit $M \gg m_H, m_Z$, the rate to $ZH$ is equal to the total rate to fermion pairs times $3/80$.} By generalizing the quarkonium result (see, e.g.,~\cite{Koller:1978qg}), we obtain
\be
\Gamma_{\BS\to ggg} = \frac{5\l(\pi^2-9\r)}{27\pi}\,\frac{A_R^2 C_R^3}{D_R}\,\alpha_s^3\bar\alpha_s^3 m
\ee
The rate for $\BS\to \gamma gg$ is described by the same expression with the replacement
\be
\frac{A_R^2}{D_R}\,\alpha_s
\to \frac{9}{20}\,D_R C_R^2\, Q^2\alpha
\ee
The rate for $\BS\to Zgg$ is given by the expression for $\gamma gg$ times $\tan^2\theta_W$.

The branching ratio to any single flavor of leptons, which with just (\ref{annih-ffbar}) would always be $12.5\%$, remains $\sim 10\%$ for $R = \vv{3}$ and $\vv{8}$ but becomes smaller for the other representations, sometimes even below $1\%$, depending on the charge, representation and mass. The resulting $\ell^+\ell^-$ signal (for any single flavor of leptons) is shown in figure~\ref{fig-cs-dilepton}, which also shows our estimates of the exclusion limits from ATLAS~\cite{ATLAS-CONF-2012-007} and CMS~\cite{Chatrchyan:2012it}.\footnote{We plot the ATLAS limit on a narrow-width RS graviton ($k/\bar M_{\rm Pl} = 0.1$) and the CMS limit on a $Z'_{\rm SSM}$ (whose width is $\sim 3\%$). In both cases we use the combined limit from $e^+e^-$ and $\mu^+\mu^-$ channels.}

\subsubsection{The cases of spin-0 and spin-1 particles}

Bound states of spin-0 or spin-1 particles are very unlikely to have a dilepton signal comparable to that of spin-$\frac12$ particles.

Spin-0 particles cannot form S-wave spin-1 bound states, while the color-singlet S-wave spin-1 bound state of spin-1 particles has the quantum numbers $J^{PC} = 1^{+-}$ so it cannot decay to $\ell^+\ell^-$ (nor can it be easily produced). As a result, the $\ell^+\ell^-$ signal, primarily from spin-1 P waves, is likely to be several orders of magnitude smaller than for spin-$\frac12$ particles.  First, the P-wave states will preferentially transition into S waves rather than annihilate to leptons (or anything else), in part because their annihilation rates are suppressed by the vanishing wavefunction at the origin.   Second, the P waves have no large production modes.  Their direct production cross section will be suppressed (relative to the S waves of the $j=\frac12$ case) because of their vanishing wavefunction at the origin. Meanwhile, production of P-wave states via a radiative transition is also suppressed.  Radiation from a D wave is suppressed because the direct production of D waves is very rare, as the derivative of their wavefunction also vanishes at the origin.  Meanwhile, although excited S waves are easier to produce and may also transition to the P waves, their annihilation rates are large, making radiative transitions rare.  In summary, we do not expect to be able to see $\ell^+\ell^-$ resonances if $j=0$ or 1.

\section{Widths}
\label{sec-widths}

Earlier, in section~\ref{sec-BS}, we considered the splittings between the various states relative to the detector resolution.  Here we will briefly consider the effects of the states' intrinsic widths. It is important to check that the widths $\Gamma$ of the bound states are smaller than the binding energies $E_b$, since otherwise the bound state's lifetime would be short compared to its orbital time, and rather than appearing as a distinct resonance in, for example, the $\gamma\gamma$ spectrum, it would merely distort the continuum enhancement of $\gamma\gamma$ production by $X\bar X$ loops.  It is also important experimentally to know whether the states are narrower or wider than the intrinsic resolution of the detectors.  Besides the possibility that the constituents are short-lived (which we will not discuss here since the lifetime is very model-dependent), there is an irreducible contribution to the bound state width from the annihilation processes, which can be significant for bound states of particles in high color representations.

The expressions for the widths of all the relevant bound states are derived in appendix~\ref{app-widths}. We find that  $\Gamma/|E_b| \ll 1$ in all cases that we considered in the previous section, so we indeed have well-defined bound states below threshold.

Now let's turn to the experimental resolutions. In the $\gamma\gamma$ channel, where the signal is coming from ${\cal R}=\vv{1}$, $J=0,2$ bound states and the resolution in invariant mass is $\sim 1\%$, the widths are negligible for the low representations, but become comparable to the resolution for $R=\vv{10}$ and $\vv{15}$. In the $\gamma$+jet channel, where the signal is coming from ${\cal R}=\vv{8}$,  $J=0,2$ bound states and the experimental resolution is $\sim 3\%$, the widths are negligible. In the dijet channel, where the signal is coming from all the possible representations $\cal R$ and the resolution is $\sim 5\%$, the widths are again negligible. In the $\ell^+\ell^-$ channel, where the signal is coming from ${\cal R}=\vv{1}$, $J=1$ bound states (a fraction of which are produced from P waves), the widths are negligible relative to the resolution, which is $\sim 1\%$ for $e^+e^-$ and larger for $\mu^+\mu^-$.

In many cases the bound state is long-lived on the QCD scale, $\Gamma \lesssim \Lambda_{\rm QCD}$. If this happens to colored bound states, they will hadronize with light quarks and gluons before annihilating. However, the $X\bar X$ core of the state, which is generally much smaller than the QCD length scale, will not be much affected by this ``brown muck'', whose size is of order $1/\Lambda_{\rm QCD}$, and in particular our wave-function and annihilation calculations will be largely unaffected.  There is one exception to this statement.   For exceptionally long bound-state lifetimes, namely when $\Gamma \lesssim \Lambda_{\rm QCD}^2/m$, and if the bound state has spin, its polarization may start oscillating if the resulting hadron is not in its total spin eigenstate, as has been discussed recently in more detail in~\cite{Grossman:2008qh}. This effect would change the angular distribution of the annihilation products. However, since this can only happen to colored bound states, the angular distribution (\ref{angular-J2}) in the $\gamma\gamma$ channel is always unaffected. Moreover, a simple comparison of the annihilation rates to $\Lambda_{\rm QCD}^2/m$ indicates that the dijet and $\gamma$+jet channels are unlikely to be affected even for the lowest masses that we considered in this paper.

\section{Strategy}
\label{sec-strategy}

In this section we will show how the bound state signals can be used for determining the various quantum numbers of the new particles. As discussed in section~\ref{sec-QN}, unstable particles $X$ must have integral charges for $R = \vv{8}$ and $\vv{10}$ and fractional charges which are multiples of~$\frac{1}{3}$ for $R=\vv{3}$, $\vv{6}$ and $\vv{15}$. We will appeal to this fact when needed. In our general discussion we will assume the particle to be charged and will dedicate a separate subsection to neutral particles, for which only the dijet signal is present.  We will also devote a subsection to representations higher than the $\vv{10}$ and $\vv{15}$, for which some qualitative and semi-quantitative statements are merited despite the lack of any trustworthy quantitative technique.

\subsection{Determining spin}
\label{sec-spin}

One can identify spin-1 particles based on the angular distribution in the diphoton (or photon+jet) channel. For bound states of spin-0 or spin-$\frac12$ particles it will be isotropic, while for spin-1 particles the dominant contribution will be coming from $J=2$ bound states with the angular distribution (\ref{angular-J2}).

For distinguishing between spin-0 and spin-$\frac12$ particles, three methods are available. One striking indication that the constituent particles have spin $\frac{1}{2}$ would be the presence of an $\ell^+\ell^-$ signal. This is a good method as long as the charge and the color representation are not too small for the observation of this signal within a reasonable amount of time.

Another method is to use the angular distribution in the dijet channel. For spin-0 particles, the distribution would be isotropic, while for spin-$\frac12$ particles the contribution of the $J=1$ bound states produced in the $q\bar q$ channel (which exists for $R \neq \vv{3}$) would follow (\ref{angular-J1}). This method is useful as long as the $q\bar q$ channel contribution to the cross section is comparable to or larger than the $gg$ contribution. This condition is satisfied for $m \gtrsim 500$~GeV in the case of the 7~TeV LHC and $m \gtrsim 1000$~GeV in the case of the 14~TeV LHC.

The third method is to use a precise measurement of the cross section of one of the signals. In the dijet channel, the cross sections for bound states of spin-$\frac12$ particles are more than twice as large as those of spin-0 particles in the same representation $R$. In the photon+jet and diphoton channels, they are exactly twice as large, for equal $R$ and $Q$. (Since there is only a discrete number of possibilities for $R$ and $Q$, the fact that the signal depends on these variables as well, to be explored in more detail in the next subsection, will not generically prevent us from determining the spin.) However, these differences are not large compared with the uncertainty of our results, so this method will only become viable once a sufficiently precise higher-order calculation is available.

\subsection{Determining charge and color representation}

\begin{figure}[t]
\begin{center}
\includegraphics[width=0.485\textwidth]{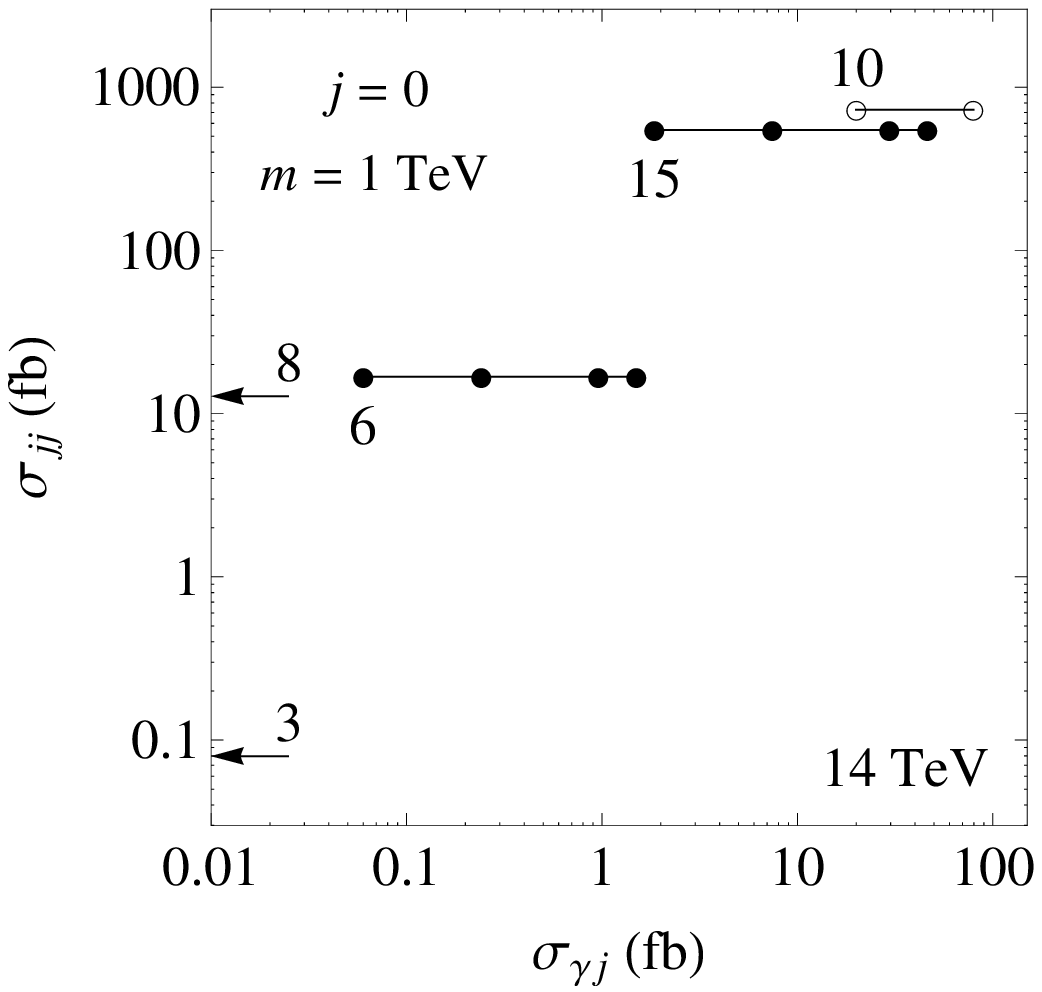}\q
\includegraphics[width=0.48\textwidth]{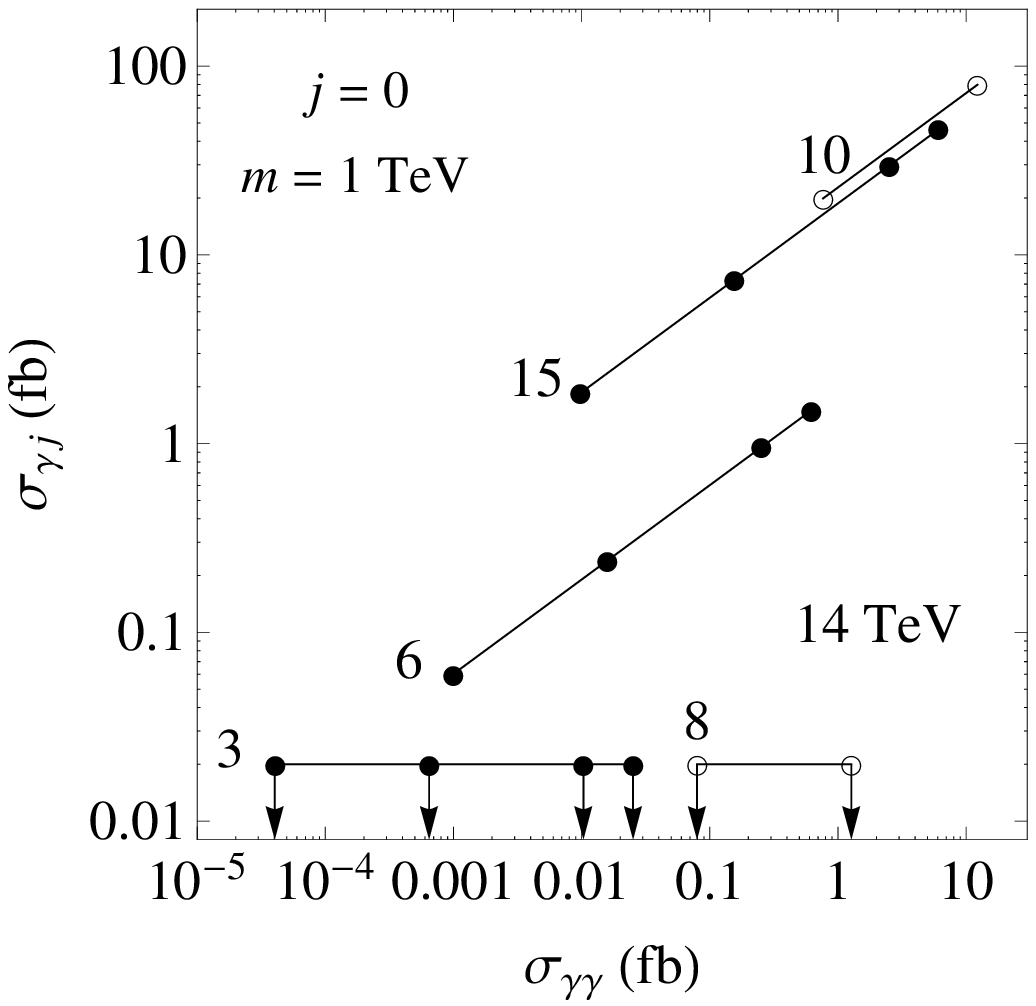}
\caption{Cross sections for bound states of spin-0 particles in the dijet versus $\gamma j$ channel (left), and in the $\gamma j$ versus diphoton channel (right).  The right plot can be used also for spin-$\frac12$ particles after multiplying all the signals by 2; either plot can be used for spin-1 particles after multiplying by 19. For $\vv{3}$, $\vv{6}$ and $\vv{15}$ each line contains four full circles corresponding to $Q = \frac{1}{3},\frac{2}{3},\frac{4}{3},\frac{5}{3}$ (from left to right) and for $\vv{8}$ and $\vv{10}$ two empty circles corresponding to $Q = 1,2$ (from left to right). Representations $\vv{3}$ and $\vv{8}$ do not have a $\gamma j$ signal. The plots refer to the 14~TeV LHC and assume $m = 1$~TeV.}
\label{fig-cs-combination-j0}
\end{center}
\end{figure}

\begin{figure}[t]
\begin{center}
\includegraphics[width=0.485\textwidth]{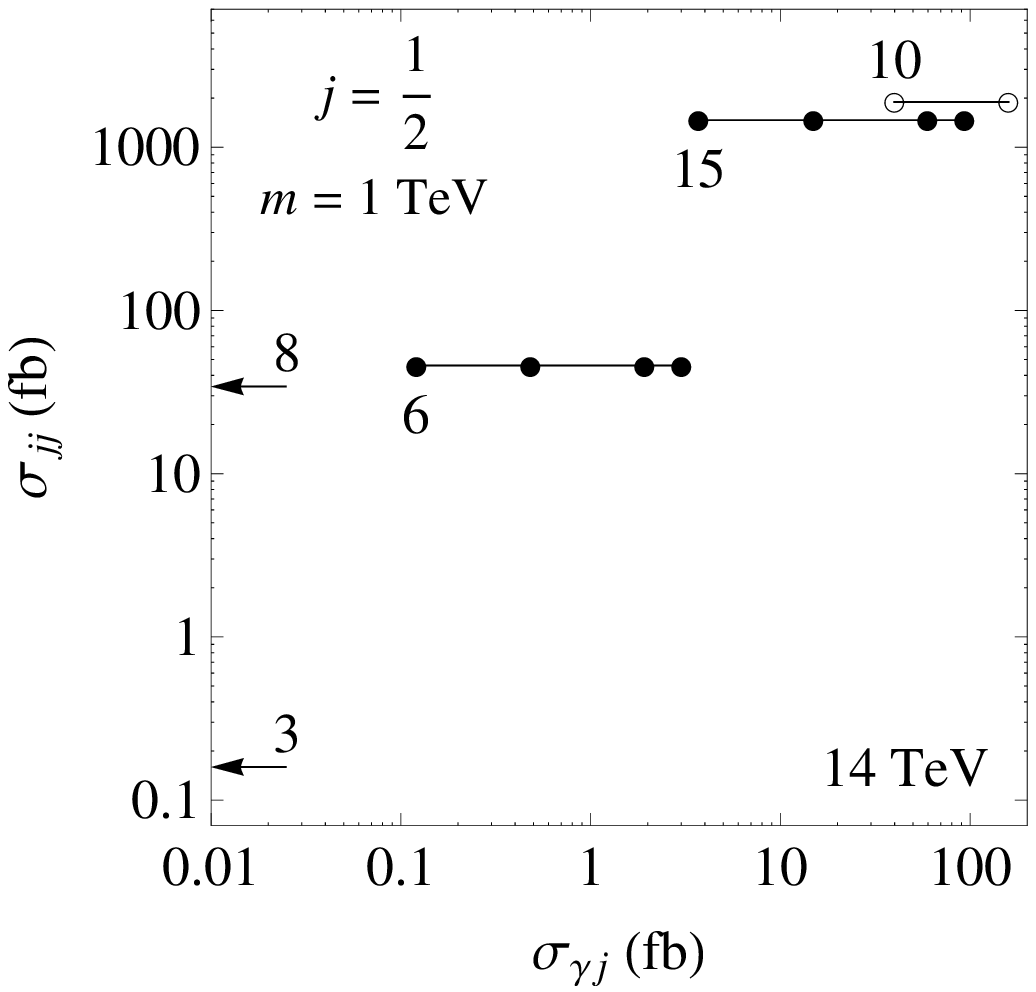}\q
\includegraphics[width=0.485\textwidth]{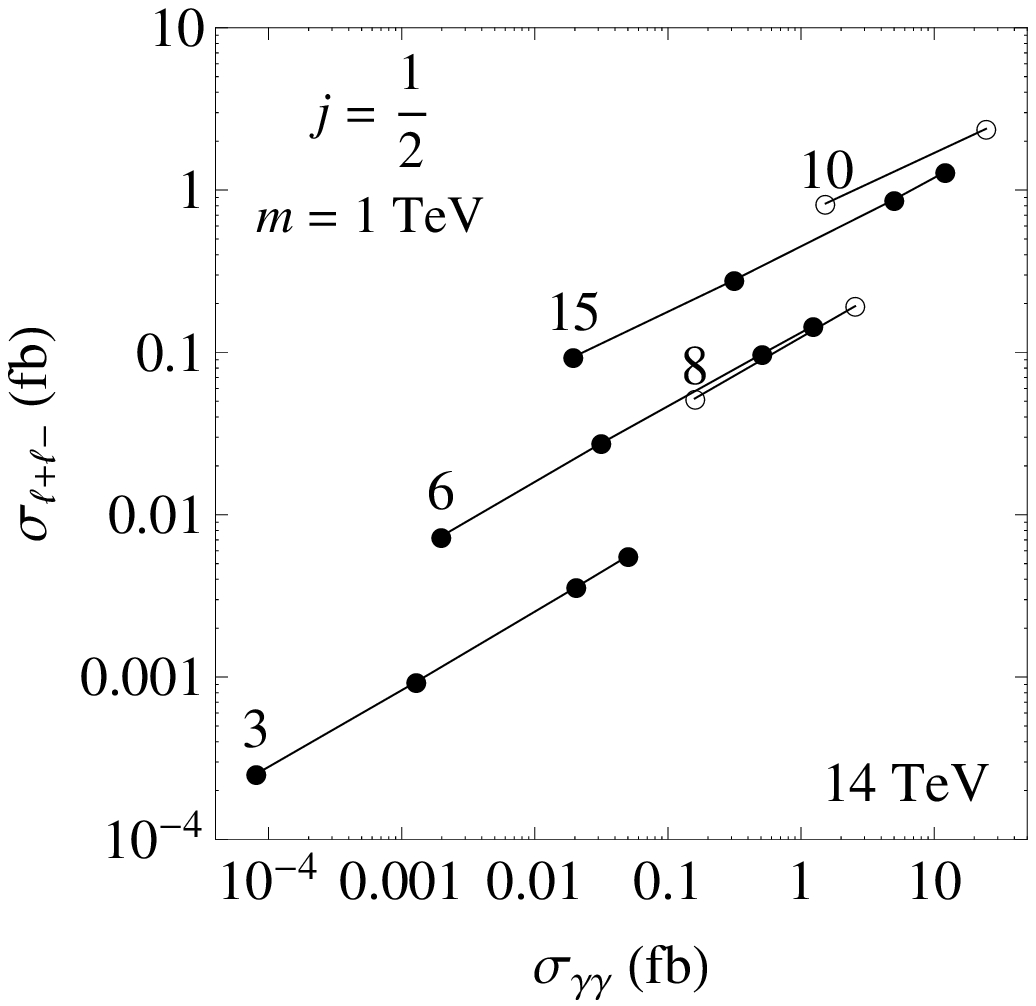}
\caption{Cross sections for bound states of spin-$\frac12$ particles in the dijet versus $\gamma j$ channel (left), and in the dilepton versus diphoton channel (right).}
\label{fig-cs-combination-j12}
\end{center}
\end{figure}

The cross sections in the dijet and $\gamma$+jet channels are likely to be measured first and they alone will be sufficient for determining the representation and the charge in many cases, as demonstrated in figure~\ref{fig-cs-combination-j0}~(left) for bound states of spin-0 or spin-1 particles and in figure~\ref{fig-cs-combination-j12}~(left) for bound states of spin-$\frac12$ particles. It will still be impossible to know the charge of particles in the $\vv{3}$ and $\vv{8}$ representations, which do not have a $\gamma$+jet signal, and some ambiguity may exist in some of the other cases due to the unknown $K$-factors. Most of these issues will be resolved once the cross section in the diphoton channel is measured too, as demonstrated in figure~\ref{fig-cs-combination-j0}~(right). In the case of spin-$\frac12$ particles, the dilepton channel can be used as well, as shown in figure~\ref{fig-cs-combination-j12}~(right).

\subsection{Determining mass}

The mass $m$ of the constituent particles is immediately known to within a few percent from the location of the resonance peak (since $M \approx 2m$). An even more accurate measurement of the mass is possible by using $M = 2m + E_b$ since the binding energy $E_b$ is calculable once $j$, $R$ and $Q$ are all known (although the uncertainty in leading-order expressions such as~(\ref{Eb-wf}) is large and a higher order calculation is essential).

\subsection{More general scenarios}

Note that if all signals are present, the problem we have posed is over-determined.  The full set of quantum numbers can be extracted from the cross sections in the $\gamma$+jet and diphoton channels (see figure~\ref{fig-cs-combination-j0}, right), the angular distribution in one of those channels which is anisotropic for $j=1$, and the observation of a dilepton signal which may be needed to break a degeneracy between $j=0$ and $j=\frac{1}{2}$.  This leaves several other observables, including the cross section in the dijet channel, certain angular distributions of the annihilation products, and (for spin-$\frac{1}{2}$ particles) the cross section and $p_T$ distribution of the dilepton resonance.  One may also try to measure the $t\bar t$ signal, which is available via (\ref{xsec-qqbar-dijet}) for spin-$\frac12$ particles (we have not discussed this channel in detail because we found that with currently available searches it is inferior to the dijet channel).  Besides that, we have not utilized the signal shape, i.e., the resonance widths and the mass differences between the various resonances contributing to a given scenario, which for a few cases (see sections~\ref{sec-BS} and~\ref{sec-widths}) may be larger than the experimental resolution in the diphoton and dielectron channels.

These extra observables may be used to relax some of our assumptions. For example, if the decay rate of the constituent particles is non-negligible relative to the annihilation rate, all the signals will be scaled down by a fixed branching ratio, and the dijet signal provides the additional measurement that can be used for determining it.  The additional observables may be helpful in resolving ambiguities introduced by the possibility of large $K$-factors, or in testing whether the particles have other large couplings besides Standard Model gauge interactions (such as discussed for spin-1 particles in appendix~\ref{app-spin1}).  And they allow us to generalize our strategy to include cases where $X$ is part of a roughly degenerate multiplet, either of electroweak $SU(2)$ or of some as yet unknown approximate global symmetry.  For $SU(2)$ multiplets, our results for the $\gamma\gamma$, $\gamma$+jet and dijet signals would still apply for each member of the multiplet separately, while determining the dilepton signal would require small modifications to our analysis.

\subsection{The case of neutral particles}

\begin{figure}[t]
\begin{center}
\includegraphics[width=0.7\textwidth]{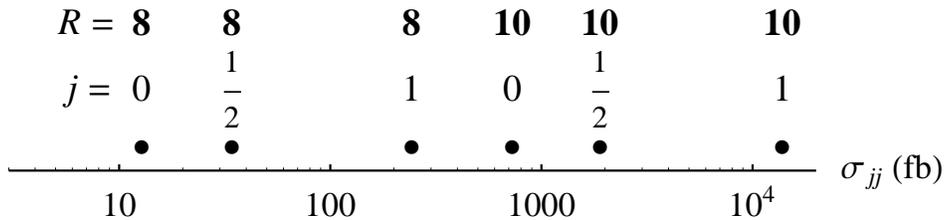}
\caption{Dijet signal cross sections for particles that may be neutral (i.e., $R = \vv{8}$ or $\vv{10}$) with spin $j = 0$, $\frac12$, $1$ for $m = 1$~TeV at the 14~TeV LHC.}
\label{fig-cs-dijet-neutral}
\end{center}
\end{figure}

For $R=\vv{8}$ and $\vv{10}$, the particles may be neutral, and then only the dijet signal will be present. In principle, as demonstrated in figure~\ref{fig-cs-dijet-neutral}, the size of the signal allows one to determine the representation and the spin. In cases where the cross sections differ by not much more than a factor of $2$ some ambiguity may remain due to unknown $K$-factors before a higher order calculation is available. In such situations one can use the angular distribution to obtain further information. We have already discussed this in section~\ref{sec-spin} for $j=0$ \emph{vs.} $j=\frac12$. These cases can also be distinguished from $j=1$ which will be dominated by the angular distribution for the $J=2$ state [which will be the same as in~(\ref{angular-J2})], with only a small isotropic component.

\subsection{Even higher color representations}
\label{sec-high-reps-results}

Following the discussion in section~\ref{sec-higher-order-intro}, we have been restricting our attention to the first five color representations from table~\ref{tab-reps} since for the $\vv{27}$, $\vv{24}$, $\vv{15'}$ (and higher representations) our leading-order methods for computing the bound states are inadequate. Still, such particles may in principle exist in nature and give rise to bound state annihilation signals despite the fact that we cannot compute them precisely. Let us therefore describe the leading-order predictions for such signals anyway, at least to get an idea about some of their qualitative features.

The signals for the $\vv{27}$, $\vv{24}$ and $\vv{15'}$ are typically significantly larger than those of the $\vv{15}$ and $\vv{10}$ (for same spin and charge) due to their larger color factors. The $\vv{27}$ has vanishing anomaly coefficient and therefore, like the $\vv{8}$, does not have a $\gamma$+jet signal at the leading order.

One qualitatively new feature of these high representations is that the annihilation rates for the color-singlet $J=0$ and $J=2$ bound states (which would contribute to the diphoton and dijet resonances) are so large that the bound state widths are comparable to or even somewhat larger than the binding energy, especially for $j = \frac12$ or $1$. As a result, the signatures of these states will be much less pronounced. However, their radial excitations (see eq.~(\ref{zeta})) would still appear as narrow peaks since their annihilation rates are suppressed by $1/n^3$. In the dijet channel, all the colored bound states are still sufficiently narrow so their contributions remain intact. As a result, even without the overly broad states, the diphoton and dijet signals for $R = \vv{27}$, $\vv{24}$, and $\vv{15'}$ are still somewhat larger than those of the $\vv{15}$ and $\vv{10}$. We also find for these representations that in some cases the bound state widths become comparable to (but not much larger than) the experimental resolution in various channels (for details, see appendix~\ref{app-widths}).

The determination of quantum numbers proceeds along the same general lines as we have discussed for the lower representations. A difficulty arises though in distinguishing between the $\vv{24}$ and $\vv{15'}$ because their color factors are similar and the allowed charges are fractional in both cases. For $j=0$ or $1$, information from pair production above the threshold can help. According to table~\ref{tab-Njets}, the $\vv{15'}$ would decay to at least 4 jets, while for the $\vv{24}$, decays to 3 jets are possible.

Finally, it is interesting to note that the binding energies for the $\vv{27}$, $\vv{24}$, and $\vv{15'}$ become large compared to the bound state mass, as large as $\sim 0.1 M$ for S-wave color-singlet bound states. As a result, the splittings between some of the states may exceed the detector resolution so that several distinct peaks will be observable. The dielectron channel (available in the $j=\frac12$ case) seems to be especially promising, while the diphoton channel (available for any $j$) is somewhat less so because the ground states of the bound states contributing to that channel are very broad, as we have just discussed. Unfortunately, it will not be straightforward to extract much quantitative information from the observed splittings, since it will be difficult to compute properties of bound states of the $\vv{27}$, $\vv{24}$, and $\vv{15'}$ reliably.

\section{Summary and Discussion}
\label{sec-summary}

Bound state annihilation signals provide a uniquely powerful and largely model-independent probe that may be utilized at the LHC to study new colored pair-produced particles.  One may use them to investigate hadronically-decaying particles using photons and leptons, measure precisely and unambiguously the masses of particles with partly-invisible decays, and in especially difficult cases even exclude or discover particles more effectively than is possible with more direct methods.

Specifically, we have argued that if new colored and charged pair-produced particles are present, then as long as their decays are not too rapid, their bound states will generically give rise to dijet, $\gamma$+jet, $\gamma\gamma$, and sometimes $\ell^+\ell^-$ resonances. We have computed the signals in the various channels and demonstrated that they typically allow unambiguous determination of the constituent particles' spin, color representation, and electric charge. In fact, only a subset of the signals is needed.  There are more observables than unknowns, so many of our assumptions can be relaxed, extending the framework to multiplets of electroweak-$SU(2)$ or other as-yet unknown symmetries, and to situations in which the constituent decays compete with the bound state annihilation rates.

The non-observation of new resonances due to bound state annihilation imposes constraints on models containing new pair-produced colored particles. For large representations, the bound state cross sections (dominated by the dijet channel, figure~\ref{fig-cs-dijet}) are very large, sometimes of the order of the pair production cross sections (figure~\ref{fig-cs-pairs-LHC}). This is perhaps surprising, but easy to understand. First, the usual suppression of bound state formation by $|\psi(\vv{0})|^2 \propto C^3\bar\alpha_s^3$ is much reduced for higher representations. Second, the parton luminosities fall quickly with energy, and the bound states, which are slightly below the threshold, benefit from much higher luminosities than the free pairs produced well above the threshold. Happily, the model dependence of the bound state signals is usually orthogonal to that of the pair production and decay signatures. The limits from bound states depend only on a relatively mild condition --- that the particle's decay rate is not too fast --- and not at all on its decay modes. As a result, in cases with obscure decays the bound state searches can be competitive with, and thereby complementary to, direct searches.

For example, with the full 7~TeV LHC dataset, the mass limit on pair-produced fermions in the $\vv{10}$ representation (which would decay to 3 quarks) is only 610~GeV from the direct search (see figure~\ref{fig-cs-pairs-LHC}), but 780~GeV from the dijet resonance search (figure~\ref{fig-cs-dijet}). If these particles are charged with $Q=2$, all the resonances that we studied set better limits, with the strongest limit of 860~GeV coming from the diphoton channel (figure~\ref{fig-cs-diphoton}). Color-octet fermions with $Q=2$ (which would also decay to 3 quarks) are excluded up to 650~GeV by the diphoton search, while the limit from the direct search is only 500~GeV. For color-sextet fermions with $Q=4/3$ or $5/3$, the diphoton channel sets a better limit as well (550 or 590~GeV, respectively, \emph{vs}. 500~GeV).

We have been describing the bound states, as well as the short-distance parts of their production and annihilation processes, at the leading order in $\alpha_s$ (although we did distinguish between $\alpha_s$ at the scale of the hard process and $\bar\alpha_s$ at the scale of the Bohr radius). This is approximately valid in the limit that $C\bar\alpha_s$ is small. For the range of masses we considered, higher-order corrections can be large ($K$-factors of $\sim 2$ or even larger for production, large corrections to annihilation rates, and substantial corrections to the bound states themselves) and will need to be computed, especially if such signals are observed. Note however that since the possible values for the spin, color representation and charge are discrete, our methods for determining the quantum numbers will still be useful even in the presence of large theoretical uncertainties. Also, since in most of the known cases the corrections are positive, the reach of the bound state signals will probably be greater than what we have presented.

Since the existence of new colored and charged particles is a very generic possibility, and bound state signals are substantially model-independent and appear even for particles that decay in obscure ways, we suggest including the bound state signals as benchmark models for resonance searches. In particular, regardless of the original motivation for a particular resonance search, there is clearly value in presenting model-independent limits on the cross sections of narrow resonances. Some of our examples emphasize the need to keep improving the exclusion limits even at relatively low masses. In some cases this will require analyzing data in regimes where the triggers are not fully efficient or collecting data with prescaled triggers.  We hope this work will help motivate such efforts.

\section*{Acknowledgments}

We would like to thank Bogdan Dobrescu, Yuri Gershtein, Yuval Grossman, Eva Halkiadakis, Maxim Perelstein, Aaron Pierce, Gavin Salam, Torbj\"{o}rn Sj\"{o}strand, and Kai Yi for discussions and communications related to various aspects of this project. The research of YK and MJS is supported in part by DOE grant DE-FG02-96ER40959.  The research of MJS is also supported in part by NSF grant PHY-0904069.

\appendix

\section{Group theory identities for $SU(N)$}
\label{app-group-theory}

In the following summary of our conventions and useful identities, $T_R^a$ denote the generators in representation $R$ of $SU(N)$, $D_R$ is the dimension of the representation, $C_R$ is the quadratic Casimir, $T_R$ is the index, and $A_R$ is the anomaly coefficient (which are listed for various representations of $SU(3)$ in table~\ref{tab-reps}):
\be
[T_R^a,T_R^b] = if^{abc}\,T_R^c \;, \qq
\l(T_R^a\r)_{ij}\l(T_R^a\r)_{jl} = C_R\, \delta_{il}
\ee
\be
\mbox{Tr}\, T_R^a = 0 \;, \q
\mbox{Tr}\l(T_R^a T_R^b\r) = T_R\, \delta^{ab} \;, \q
T_R = \frac{D_R}{D_A} C_R
\ee
\be
\mbox{Tr}\l(T_R^a\l[T_R^b,T_R^c\r]\r) = i f^{abc}\, T_R \;, \q
\mbox{Tr}\l(T_R^a\l\{T_R^b,T_R^c\r\}\r) = \frac{1}{2} d^{abc} A_R
\ee
\be
f^{abc}f^{abd} = N\delta^{cd} \;,\q
d^{abc}d^{abd} = \frac{N^2-4}{N}\,\delta^{cd} \;,\q
d^{aac} = 0
\ee
\be
f^{abe}f^{ecd} + f^{cbe}f^{aed} + f^{dbe}f^{ace} = 0 \;, \qq
d^{ecd}f^{abe} + d^{aed}f^{cbe} + d^{ace}f^{dbe} = 0
\ee
\be
d^{abc}f^{adg}f^{beg} = \frac{N}{2}\,d^{cde} \;, \qq
f^{abc}f^{adg}f^{beg} = \frac{N}{2}\,f^{cde}
\ee
\be
d^{abe}d^{cde} = \frac13\l(\delta^{ac}\delta^{bd} + \delta^{ad}\delta^{bc} - \delta^{ab}\delta^{cd} + f^{ace}f^{bde} + f^{ade}f^{bce}\r)
\ee
For any representation $R = \l(a,b\r)$ of $SU(3)$,
\be
D_R = \frac12\l(a+1\r)\l(b+1\r)\l(a+b+2\r)\,,\q
C_R = \frac13\l(a^2 + 3a + ab + 3b + b^2\r)
\ee
and the triality is
\be
t_R = (a + 2b)\,\mbox{mod}\,3
\label{triality}
\ee

\section{Interactions of spin-1 particles}
\label{app-spin1}

For spin-$1$ particles $X^\mu$, gauge invariance alone does not fix the interactions with the gauge bosons completely, and we must specify the effective Lagrangian:
\be
\cL = -\frac12X_{\mu\nu}^\ast X^{\mu\nu} - i g_s X_\mu^\ast T_R^a X_\nu G^{\mu\nu a} - i e\, Q\, X_\mu^\ast X_\nu F^{\mu\nu} + m^2 X_\mu^\ast X^\mu
\label{spin-1-Lagrangian}
\ee
where $T^a_R$ is the $SU(3)$ generator in the representation $R$, $X_{\mu\nu} = D_\mu X_\nu - D_\nu X_\mu$ with $D_\mu = \pd_\mu - ig_s T_R^a A_\mu^a - ieQ A_\mu$, $G^a_{\mu\nu}$ is the gluon field strength, and $F^{\mu\nu}$ is the photon field strength. The coefficients of the second and third terms in (\ref{spin-1-Lagrangian}) were chosen such that the theory preserves tree-level unitarity at high energy (a special case being when $X^\mu$ is a gauge boson of an extended gauge group)~\cite{Borisov:1986ev}. The Feynman rules are similar to those of vector leptoquarks (with vanishing anomalous couplings)~\cite{Blumlein:1996qp,Blumlein:1996gb}.

Another subtlety is in the interaction with quarks. Pairs of particles of any spin can be produced from $q\bar q$ through a diagram with an $s$-channel gluon. However, as mentioned in appendix~\ref{app-pair-xsec}, for spin-1 particles this diagram violates unitarity at high energies, so any consistent theory must contain additional $q\bar q$ diagrams involving some new interactions. For instance there may be diagrams with the exchange of some new particles in the $t$ channel.

Therefore, for spin-1 particles, we have neglected the $q\bar q$ channel contribution to the pair production cross sections in figures~\ref{fig-cs-pairs-Tevatron}--\ref{fig-cs-pairs-LHC} because it is model-dependent. We have checked that even if we used the diverging diagram (whose eventual contribution is finite because of the falling PDFs), the cross sections for vectors in the range of masses presented in figures~\ref{fig-cs-pairs-Tevatron}--\ref{fig-cs-pairs-LHC} would still be dominated by the $gg$ channel, except in the case of particles in the $\vv{3}$ representation, where the contributions of the $gg$ and $q\bar q$ channels may be comparable.

Now let's discuss the possible effects on bound states. The unitarity-violating $q\bar q$ diagram would not produce S-wave bound states by itself because it vanishes at the threshold. However, the additional interactions that must be present in the theory may affect bound state processes.  As an example, bound states of KK gluons in a theory of universal extra dimensions, studied in~\cite{Kahawala:2011pc}, are affected by $q\bar q$ diagrams with KK quarks, which contribute to the production and annihilation of $J=2$ bound states. By examining that example (and generalizing it to particles in other color representations) we notice that the effects of the $q\bar q$ channel will typically be small (although sometimes not completely negligible) relative to the $gg$ channel as long as the KK quarks are at least twice as heavy as the KK gluons.  This suggests that in typical models there are regimes of parameter space where the bound state processes are dominated by the $gg$ channel, and this motivates us to assume, for simplicity and model-independence, that the contributions to bound states of spin-1 particles from the $q\bar q$ channel are small.

\section{Pair production cross sections}
\label{app-pair-xsec}

This appendix contains the expressions for the $X\bar X$ pair production cross sections, relevant to figures~\ref{fig-cs-pairs-Tevatron}--\ref{fig-cs-pairs-LHC} (where $\alpha_s$ and the NLO PDFs~\cite{Martin:2009iq} were evaluated at the scale $m$).

For scalars~\cite{Borisov:1986ev,Chivukula:1991zk,Chen:2008hh,DelNobile:2009st}:
\bea
\hat\sigma(gg\to X\bar X) &=& \frac{T_R}{16}\,\frac{\pi\alpha_s^2}{\hat s}\l[4C_R\l(2-\beta^2\r)\beta + \l(3-5\beta^2\r)\beta \frac{}{}\r.\nn\\
&&\qquad\qquad\l. - \l(2C_R(1-\beta^4) + 3\l(1-\beta^2\r)^2\r)\ln\frac{1+\beta}{1-\beta}\r] \\
\hat\sigma(q\bar q \to X\bar X) &=& \frac{4}{27} T_R \frac{\pi\alpha_s^2}{\hat s}\beta^3
\eea
where $\beta = \sqrt{1-4m^2/\hat s}$. These expressions agree with those for squarks~\cite{Beenakker:1996ch} for $R = \vv{3}$.\footnote{The result for scalars in the $gg$ channel in~\cite{Borisov:1986ev} is incorrect. It does not reduce to the expression for squarks. Ref.~\cite{Blumlein:1996qp} has also found that they disagree with the result of~\cite{Borisov:1986ev} for scalars.}

For fermions~\cite{Borisov:1986ev,Chivukula:1990di,DelNobile:2009st}:
\bea
\hat\sigma(gg\to X\bar X) &=& \frac{T_R}{8}\,\frac{\pi\alpha_s^2}{\hat s}\l[ - 4C_R\l(2-\beta^2\r)\beta - \l(9 - 5\beta^2\r)\beta \frac{}{}\r.\nn\\
&&\qquad\qquad\l. + \l(2 C_R\l(3 - \beta^4\r) + 3\l(1-\beta^2\r)^2\r) \ln\frac{1+\beta}{1-\beta}\r] \\
\hat\sigma(q\bar q \to X\bar X) &=& \frac{8}{27}T_R\frac{\pi\alpha_s^2}{\hat s}\beta\l(3-\beta^2\r)
\eea
These results agree with those for heavy quarks~\cite{Bonciani:1998vc} for $R = \vv{3}$.

For vectors~\cite{Borisov:1986ev}:
\bea
\hat\sigma(gg\to X\bar X) &=& \frac{T_R}{64}\,\frac{\pi\alpha_s^2}{m^2}\l[4C_R\l(22 - 9\beta^2 + 3\beta^4\r)\beta + 3\l(19 - 24\beta^2 + 5\beta^4\r)\beta \frac{}{}\r. \nn\\
&&\l.\qqqq\; - \frac{12m^2}{\hat s}\l(2C_R\l(1-\beta^4\r) + 19 - 6\beta^2 + 3\beta^4\r) \ln\frac{1+\beta}{1-\beta}\r]
\label{gg>GG*} \\
\hat\sigma(q\bar q\to X\bar X) &=& \frac{T_R}{108}\,\pi\alpha_s^2\,\frac{\hat s}{m^4}\,\l(27-26\beta^2+3\beta^4\r)\beta^3
\label{qqbar>GG*}
\eea
Note that in the $gg$ channel in the limit $\hat s\to\infty$
\be
\hat\sigma(gg\to X\bar X) \to T_R C_R\,\frac{\pi\alpha_s^2}{m^2}
\ee
which is consistent with unitarity. For $R = \vv{3}$, eq.~(\ref{gg>GG*}) reduces to the result obtained for vector leptoquarks (with no anomalous couplings) in~\cite{Blumlein:1996qp} and for $R = \vv{8}$ to twice that of the coloron of~\cite{Dicus:1994sw,Dobrescu:2007yp} (because the coloron field is real).
On the other hand, the expression for the $q\bar q$ channel, eq.~(\ref{qqbar>GG*}), where only the $s$-channel gluon diagram has been included, diverges with $\hat s$ and thus violates unitarity, as has been already noted in~\cite{Borisov:1986ev}. A UV completion is needed in this case, and the resulting contribution from the $q\bar q$ channel will be model-dependent. We discuss this issue further in appendix~\ref{app-spin1}.

\section{Widths}
\label{app-widths}

\begin{table}[t]\small{
$$\begin{array}{|c||c|c|c|c|c|c|c|c|}\hline
{\cal R}\downarrow\; R\rightarrow &\q\;\vv{3}\;\q&\q\vv{8}\q&\q\vv{6}\q&\q\vv{15}\q&\q\vv{10}\q&\q\vv{27}\q&\q\vv{24}\q&\q\vv{15'}\q\\\hline\hline
\vv{1}        & 0.0008 &  0.02 & 0.02  & 0.2   & 0.2   & 1.3  & 1.3  & 1.1  \\\hline
\vv{8}        &    -   &       & 0.003 & 0.006 & 0.04  &      & 0.09 & 0.2  \\\hline
\vv{\tilde 8} &    -   & 0.003 &   -   & 0.03  &   -   & 0.2  & 0.1  &  -   \\\hline
\vv{27}       &    -   &   -   &   -   & 0.006 & 0.008 & 0.09 & 0.09 & 0.09 \\\hline
\end{array}$$}
\caption{Annihilation widths compared with the binding energies, $\Gamma/|E_b|$, for $j=0, J=0$. The numbers for $j=\frac12, J=0$ are larger by a factor of $2$, and for $j=1$ by a factor of $3$ for $J=0$ and $16/5$ for $J=2$. We assumed $m = 1$~TeV (although the dependence on $m$ is only through the running of $\alpha_s$). We have not specified the widths for ${\cal R}=\vv{8}$ bound states of $R=\vv{8}$ and $\vv{27}$ since they do not couple to $gg$. The numbers shown for ${\cal R}=\vv{27}$ are based on the sum (\ref{annih-27}), so for $R=\vv{15}$, $\vv{27}$, $\vv{24}$, where there is more than one way to get ${\cal R}=\vv{27}$, these are upper bounds.}
\label{tab-width-Eb}\small{
$$\begin{array}{|c||c|c|c|c|c|c|c|}\hline
{\cal R}\downarrow\; R\rightarrow &\q\vv{8}\q&\q\vv{6}\q&\q\vv{15}\q&\q\vv{10}\q&\q\vv{27}\q&\q\vv{24}\q&\q\vv{15'}\q\\\hline\hline
\vv{8}  & 0.008 & 0.008 & 0.06 & 0.05 & 0.3 & 0.3 & 0.2 \\\hline
\end{array}$$}
\caption{$\Gamma/|E_b|$ for color-octet $J=1$ bound states, relevant for $j=\frac12$.}
\label{tab-width-Eb-J1}
\end{table}

In this appendix we provide details about the bound state widths (due to the annihilation processes alone), as discussed in section~\ref{sec-widths} (and for the very high representations in section~\ref{sec-high-reps-results}).

For $J=0$ and $J=2$ bound states contributing in the $\gamma\gamma$, $\gamma$+jet and dijet channels, the dominant annihilation process is to $gg$. For constituents of a given spin $j$ and color representation $R$, the annihilation rates depend on the spin $J$ and color representation ${\cal R}$ of the bound state. We are interested in bound states with ${\cal R}=\vv{1}$, $\vv{8}$ and $\vv{27}$. For constituent particles in representations $R=\vv{8}$, $\vv{15}$, $\vv{27}$, $\vv{24}$ there are two ways to form a bound state with ${\cal R}=\vv{8}$, as indicated in table~\ref{tab-reps}, one of which is universal and described by the wavefunction~(\ref{octet-wavefunction}). In the following, we will denote it by $\vv{8}$ and the second possibility by $\vv{\tilde 8}$. Similarly, for $R = \vv{15}$, $\vv{27}$, $\vv{24}$ there are two or three kinds of ${\cal R}=\vv{27}$ bound states. For bound states of spin-$0$ particles we find
\be
\Gamma_{j=0}\l(\BS^{{\cal R}=\vv{1}}_{J=0}\to gg\r) = \frac{D_R C_R^5}{32}\,\alpha_s^2\,\bar\alpha_s^3\,m
\label{width-gg-spin0}
\ee
\be
\Gamma_{j=0}\l(\BS^{{\cal R}=\vv{8}}_{J=0}\to gg\r) = \frac{5\,A_R^2\l(C_R - \frac{3}{2}\r)^3}{192\,T_R}\,\alpha_s^2\,\bar\alpha_s^3\,m
\ee
\be
\Gamma_{j=0}\l(\BS^{{\cal R}=\vv{\tilde 8}}_{J=0}\to gg\r) = \frac1{32}\l(\frac{D_R C_R\l(C_R+\frac34\r)}{5} - \frac{5\,A_R^2}{6\,T_R}\r)\l(C_R-\frac32\r)^3\alpha_s^2\,\bar\alpha_s^3\,m
\ee
\be
\sum_{\vv{27}}\Gamma_{j=0}\l(\BS^{{\cal R}=\vv{27}}_{J=0}\to gg\r) = \frac{D_R C_R\l(C_R-\frac43\r)\l(C_R-4\r)^3}{160}\,\alpha_s^2\,\bar\alpha_s^3\,m
\label{annih-27}
\ee
where the sum in (\ref{annih-27}) is over all the possible ways to form an ${\cal R} = \vv{27}$ bound state (which is easier to obtain than computing the separate rates). For bound states of spin-$\frac12$ particles, the widths are given by the same expressions multiplied by a factor of $2$:
\be
\Gamma_{j=1/2}\l(\BS^{{\cal R}}_{J=0}\to gg\r) = 2\,\Gamma_{j=0}\l(\BS^{{\cal R}}_{J=0}\to gg\r)
\label{width-gg-spin12}
\ee
and for bound states of spin-1 particles
\be
\Gamma_{j=1}\l(\BS^{{\cal R}}_{J=0}\to gg\r) = 3\,\Gamma_{j=0}\l(\BS^{{\cal R}}_{J=0}\to gg\r)
\ee
\be
\Gamma_{j=1}\l(\BS^{{\cal R}}_{J=2}\to gg\r) = \frac{16}{5}\,\Gamma_{j=0}\l(\BS^{{\cal R}}_{J=0}\to gg\r)
\ee
Table~\ref{tab-width-Eb} compares these rates with the binding energies~(\ref{Eb-wf}), and table~\ref{tab-width-res} (top) presents the same widths as a fraction of the bound state mass $M$, which is useful for the discussion of the experimental resolution.

\begin{table}[t]\small{
$$\begin{array}{|c||c|c|c|c|c|c|c|c|}\hline
\multicolumn{9}{|c|}{j=0, J=0}\\\hline
\,{\cal R}\downarrow\; R\rightarrow &\vv{3}&\vv{8}&\vv{6}&\vv{15}&\vv{10}&\vv{27}&\vv{24}&\vv{15'}\\\hline\hline
\vv{1}        &\,3\cdot10^{-6}\,&\,3\cdot10^{-4}\,&\,4\cdot10^{-4}\,&\,9\cdot10^{-3}\,&\,1.1\cdot10^{-2}\,& (2.2\cdot10^{-2}) & (2.4\cdot10^{-2}) & (2.5\cdot10^{-2}) \\\hline
\vv{8}        & - &         -     & 2\cdot10^{-5} & 1\cdot10^{-4} & 1\cdot10^{-3} &        -        &  6\cdot10^{-3}  & 1.8\cdot10^{-2}\\\hline
\vv{\tilde 8} & - & 1\cdot10^{-5} &       -       & 7\cdot10^{-4} &     -         & 1.4\cdot10^{-2} & 1.0\cdot10^{-2} &     -          \\\hline
\vv{27}       & - &       -       &       -       & 2\cdot10^{-5} & 6\cdot10^{-5} &  2\cdot10^{-3}  &  3\cdot10^{-3}  &  4\cdot10^{-3} \\\hline
\end{array}$$\vspace{1mm}
$$\begin{array}{|c||c|c|c|c|c|c|c|c|}\hline
\multicolumn{9}{|c|}{j=\frac12, J=1}\\\hline
\,{\cal R}\downarrow\; R\rightarrow &\vv{3}&\vv{8}&\vv{6}&\vv{15}&\vv{10}&\vv{27}&\vv{24}&\vv{15'}\\\hline\hline
\vv{8} &\q - \q&\, 4\cdot10^{-5} \,&\, 5\cdot10^{-5} \,&\, 1\cdot10^{-3} \,&\, 2\cdot10^{-3} \,&\, 1.6\cdot10^{-2} \,&\, 1.7\cdot10^{-2} \,&\, 1.8\cdot10^{-2} \\\hline
\end{array}$$\vspace{1mm}
$$\begin{array}{|c||c|c|c|c|c|c|c|c|c|c|}\hline
{\cal R}=\vv{1} & \multicolumn{8}{|c|}{j=\frac12, J=1 \qqqqq}\\\hline
\,Q\downarrow\; R\rightarrow &\vv{3}&\vv{8}&\vv{6}&\vv{15}&\vv{10}&\vv{27}&\vv{24}&\vv{15'}\\\hline\hline
2/3 & 3\cdot10^{-7} & 8\cdot10^{-6} & 2\cdot10^{-5} & 1\cdot10^{-4} & 4\cdot10^{-4} & 7\cdot10^{-4} & 3\cdot10^{-3} & 6\cdot10^{-3} \\\hline
2   &\, 3\cdot10^{-6} \,&\, 7\cdot10^{-5} \,&\, 8\cdot10^{-5}\, &\, 9\cdot10^{-4} \,&\, 1\cdot10^{-3} \,&\, 7\cdot10^{-3} \,&\, 9\cdot10^{-3} \,&\, 1.3\cdot10^{-2} \,\\\hline
\end{array}$$}
\caption{Annihilation widths as a fraction of the bound state mass, $\Gamma/M$. Top table: $j=0, J=0$. Same comments as in table~\ref{tab-width-Eb} apply. For $R = \vv{27},\vv{24},\vv{15'}$ with ${\cal R} = \vv{1}$ we give the width for the first radial excitation (since the ground state is too broad). Middle table: $j=\frac12, J=1, {\cal R} = \vv{8}$. Bottom table: $j=\frac12, J=1, {\cal R} = \vv{1}$ for two values of the charge $Q$.}
\label{tab-width-res}
\end{table}

For the dijet channel in the case of $j=\frac12$ we also need the annihilation rate of $J=1$ color-octet bound states:
\be
\Gamma_{j=1/2}\l(\BS^{{\cal R}=\vv{8}}_{J=1}\to q\bar q\r) = \frac{D_R C_R\l(C_R - \frac{3}{2}\r)^3}{16}\,\alpha_s^2\,\bar\alpha_s^3\,m
\ee
This rate is compared with the binding energy in table~\ref{tab-width-Eb-J1} and with the mass in table~\ref{tab-width-res} (middle).

For the $\ell^+\ell^-$ channel, we are interested in color-singlet spin-$1$ bound states of spin-$\frac12$ particles. They cannot annihilate to $gg$ or $q\bar q$ via the strong interaction, so the width is determined by subleading processes, which are the annihilation to fermion pairs via a photon or $Z$, and the annihilation to three gauge bosons. Based on the expressions in section~\ref{sec-spin1-ann-rates}, the total rate is
\bea
&&\Gamma_{j=1/2}\l(\BS^{{\cal R}=\vv{1}}_{J=1} \to f\bar f,ggg,\gamma gg,Zgg\r) = \nn\\
&&\q\frac53\l[\frac12 D_R\, Q^2\frac{\alpha^2}{\cos^4\theta_W}
+ \frac{\pi^2-9}{9\pi}\l(\frac{A_R^2}{D_R}\,\alpha_s + \frac{9}{20}\,D_R C_R^2\, Q^2\frac{\alpha}{\cos^2\theta_W}\r)\alpha_s^2\r] C_R^3\,\bar\alpha_s^3\,m
\label{annihJ1}
\eea
These bound states are very narrow: $\Gamma/|E_b| \ll 1$ for any $R$. In the context of experimental resolution, the widths are presented as a fraction of $M$ in table~\ref{tab-width-res} (bottom).

\bibliography{bound.states}

\end{document}